\documentclass[10pt,journal,compsoc]{IEEEtran}

\usepackage[pdftex]{graphicx} 
\usepackage{booktabs}
\usepackage{times}
\usepackage[monochrome]{color}
\usepackage{courier}
\usepackage{caption}
\usepackage{listings} 
\usepackage{cite} 
\usepackage{url}  
\usepackage{graphicx}
\usepackage[cmex10]{amsmath} 
\usepackage{xfrac}
\usepackage{algorithmic} 
\usepackage[table,usenames,dvipsnames,svgnames]{xcolor} 
\usepackage{multirow}
\usepackage{flushend} 
\usepackage{subcaption}

\usepackage{wasysym}

\newcommand{\mypara}[1]{\vspace{5pt}\noindent{\textbf{#1}}}

\begin{document}

\title{A First Look at RISC-V Virtualization from an Embedded Systems Perspective}

\author{\IEEEauthorblockN{}
\IEEEauthorblockA{Centro Algoritmi - University of Minho}
\IEEEauthorblockA{\{}
}

\author{Bruno Sá,
        José Martins,
        Sandro Pinto

\par Centro ALGORITMI, Universidade do Minho, Portugal 
\par bruno.vilaca.sa@gmail.com, \{jose.martins, sandro.pinto\}@dei.uminho.pt
}


\maketitle

\begin{abstract}
This article describes the first public implementation and evaluation of the latest version of the RISC-V hypervisor extension (H-extension v0.6.1) specification in a Rocket chip core. To perform a meaningful evaluation for modern multi-core embedded and mixed-criticality systems, we have ported Bao, an open-source static partitioning hypervisor, to RISC-V. We have also extended the RISC-V platform-level interrupt controller (PLIC) to enable direct guest interrupt injection with low and deterministic latency and we have enhanced the timer infrastructure to avoid trap and emulation overheads. Experiments were carried out in FireSim, a cycle-accurate, FPGA-accelerated simulator, and the system was also successfully deployed and tested in a Zynq UltraScale+ MPSoC ZCU104. Our hardware implementation was open-sourced and is currently in use by the RISC-V community towards the ratification of the H-extension specification.
\end{abstract}



\section{Introduction}

\par In the last decade, virtualization has become a key enabling technology for servers, but also for several embedded industries such as the automotive and industrial control \cite{Heiser2011, Martins2020}. In the embedded space, the number of requirements has been steadily increasing for the past few years. At the same time, the market pressure to minimize size, weight, power, and cost (SWaP-C) has pushed for the consolidation of several subsystems, typically with distinct criticality levels, onto the same hardware platform \cite{Bechtel2019, Pinto2019}. In response, academia and industry have focused on developing hardware support to assist virtualization (e.g., Arm Virtualization Extensions), and adding upstream support for these technologies in mainstream hypervisor solutions \cite{Kloda2019, Dall2014}. 

\par Embedded software stacks are progressively targeting powerful multi-core platforms, endowed with complex memory hierarchies \cite{Burgio2017, Xu2019}. Despite the logical CPU and memory isolation provided by existing hypervisor layers, there are several challenges and difficulties in proving strong isolation, due to the reciprocal interference caused by micro-architectural resources (e.g., last-level caches, interconnects, and memory controllers) shared among virtual machines (VM) \cite{Kloda2019, Martins2020}. This issue is particularly relevant for mixed-criticality applications, where security- and safety-critical applications need to coexist along non-critical ones. In this context, a malicious VM can either implement denial-of-service (DoS) attacks by increasing their consumption of a shared resource \cite{Bechtel2019, Bechtel2020} or to indirectly access other VM's data leveraging existing timing side-channels \cite{Ge2018}. To tackle this issue, industry has been developing specific hardware technology (e.g., Intel Cache Allocation Technology) and the research community have been very active proposing techniques based on cache locking, cache/page coloring, or memory bandwidth reservations \cite{Yun2013, Mancuso2013, Kloda2019, Xu2019, Farshchi2020}. 

\par Recent advances in computing systems have brought to light an innovative computer architecture named RISC-V \cite{Asanovic2014}. RISC-V distinguishes itself from traditional platforms by offering a free and open instruction set architecture (ISA) featuring a modular and highly customizable extension scheme that allows it to scale from tiny embedded microcontrollers up to supercomputers. RISC-V is going towards mainstream adoption under the premise of disrupting the hardware industry such as Linux has disrupted the software industry. As part of the RISC-V privileged architecture specification, hardware virtualization support is specified through the hypervisor extension (H-extension) \cite{RISCV2020}. The H-extension specification is currently in version 0.6.1, and no ratification has been achieved so far. To date, the H-extension has achieved function completeness with KVM and Xvisor in QEMU. However, none hardware implementation is publicly available yet, and commercial RISC-V cores endowed with hardware virtualization support are not expected to be released in the foreseeable future.    

\par In this work, we share our experience while providing the first public hardware implementation of the latest version of the RISC-V H-extension in the Rocket core \cite{Asanovic2016}. While the specification is intended to fulfil cloud and embedded requirements, we focused our evaluation on modern multi-core embedded and mixed-criticality systems (MCS). In this context, we have ported Bao \cite{Martins2020}, a type-1, open-source static partitioning hypervisor, to RISC-V. In the spirit of leveraging the hardware-software codesign opportunity offered by RISC-V, we have also performed a set of architectural enhancements in the interrupt controller and the timer infrastructure aiming at guaranteeing determinism and improving performance, while minimizing interrupt latency and inter-hart interference. The experiments carried out in FireSim \cite{Karandikar2018}, a cycle-accurate, FPGA-accelerated simulator, and corroborated in Zynq UltraScale+ MPSoC ZCU104, demonstrate significant improvements in performance {\color{blue}($<$1\% overhead for hosted execution)} and interrupt latency {\color{blue}($>$89\% reduction for hosted execution)}, at a fraction of hardware costs {\color{blue}(11\% look-up tables and 27-29\% registers)}. We released our hardware design as open source\footnote{https://github.com/josecm/rocket-chip/tree/hyp} and the hardware is currently being used as a reference implementation by the RISC-V International\footnote{former RISC-V Foundation} to ratify the H-extension specification.  

\par In summary, with this work, we make the following contributions:

\begin{itemize} 

\item the first public and open source implementation of the latest version of the RISC-V H-extension (v0.6.1) in a Rocket Chip core (Section \ref{imp_hyp_ext});

\item a set of hardware enhancements, in the platform-level interrupt controller and the architectural timer infrastructure, to tune virtualization support for embedded and mixed-criticality requirements (Section \ref{imp_hyp_enh});

\item the port of the open source Bao hypervisor for RISC-V (Section \ref{bao_port}); 

\item the development of an open source ad-hoc testing framework that enable the raw functional validation of fine-grain features of the hypervisor specification (Section \ref{eval-fval});

\item the first public and cycle-accurate evaluation of the H-extension in a RISC-V core. We focused on hardware costs, performance overhead, inter-VM interference, and interrupt latency (Section \ref{eval});


\end{itemize}


\section{RISC-V Virtualization Support}  \label{riscv_virt_supp}

\par The RISC-V privilege architecture \cite{RISCV2020} features originally three privilege levels: (i) \textit{machine} (M) is the highest privilege mode and intended to execute firmware which should provide the supervisor binary interface (SBI); (ii) \textit{supervisor} (S) mode is the level where an operating system kernel such as Linux is intended to execute, thus managing virtual-memory leveraging a memory management unit (MMU); and (iii) \textit{user} (U) for applications. The modularity offered by the ISA allows implementations featuring only M or M/U which are most appropriate for small microcontroller implementations. However, only an implementation featuring the three M/S/U modes is useful for systems requiring virtualization. The ISA was designed from the ground-up to be classically virtualizable \cite{Popek1974} by allowing to selectively trap accesses to virtual memory management control and status registers (CSRs) as well as timeout and mode change instructions from supervisor/user to machine mode. For instance, \textit{mstatus}'s {\color{blue}trap virtual memory (TVM)} bit enables trapping of \textit{satp}, the root page table pointer, while setting the {\color{blue}trap sret (TSR)} bit will cause the trap of the \textit{sret} instruction used by supervisor to return to user mode. Furthermore, RISC-V provides fully precise exception handling, guaranteeing the ability to fully specify the instruction stream state at the time of an exception. The ISA simplicity coupled with its virtualization-friendly design allow the easy implementation of a hypervisor recurring to traditional techniques (e.g., full trap-and-emulate, shadow page tables) as well as the emulation of the hypervisor extension, described further ahead in this section, from machine mode. However, it is well-understood that such techniques incur large performance overheads.

\subsection{Hypervisor Extension}
\begin{figure}[t]
    \centering
        \includegraphics[width=0.47\textwidth,clip]{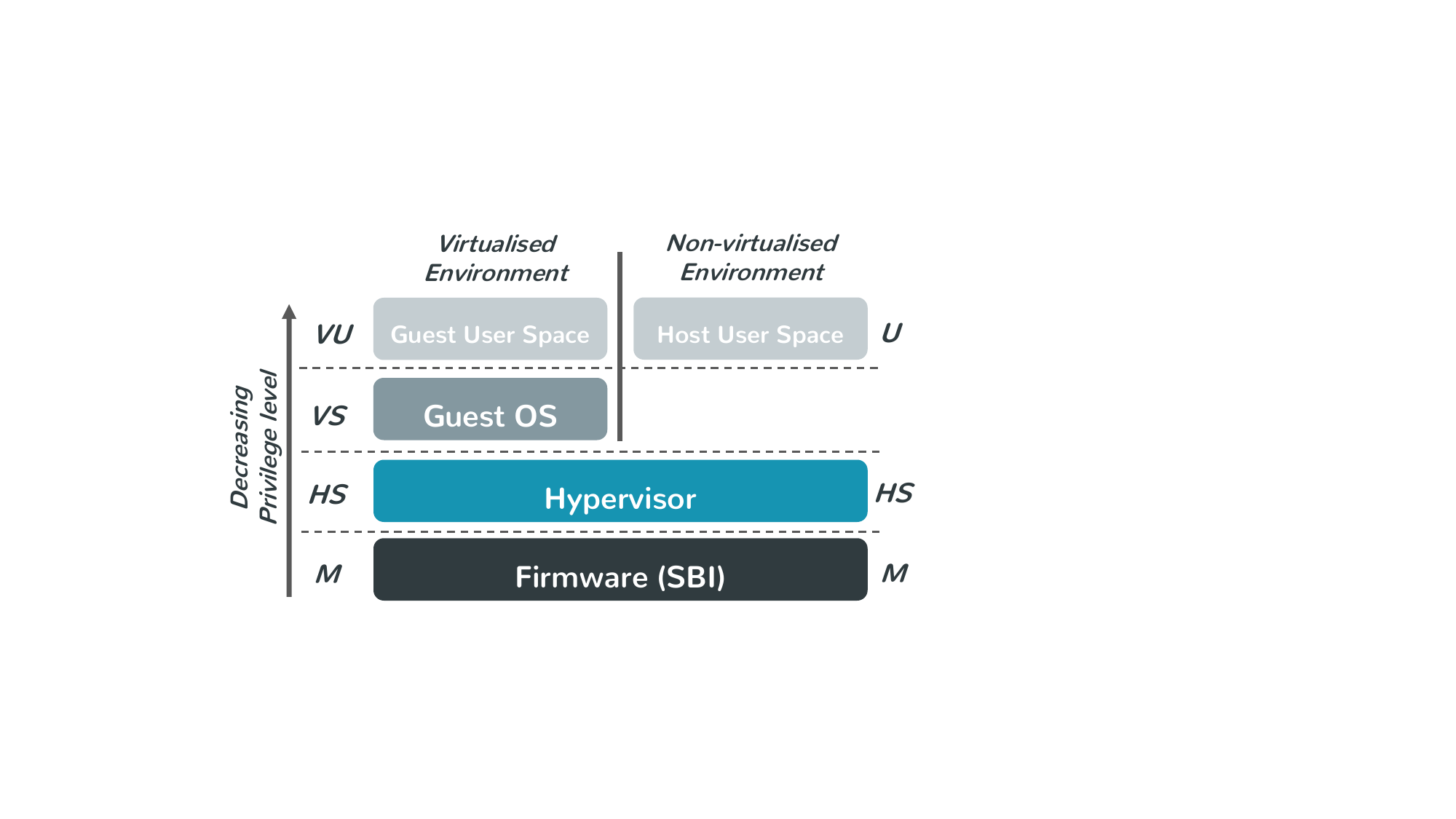}
    \caption{RISC-V privileged levels: machine (M), hypervisor-extended supervisor (HS), virtual supervisor (VS), and virtual user (VU).}
    \label{fig:exec_modes}
\end{figure}

Like most other mainstream ISAs, the latest draft of the RISC-V privilege architecture specification offers hardware virtualization support, i.e., the optional hypervisor extension ("H"), to increase virtualization efficiency. As illustrated by Fig. \ref{fig:exec_modes}, the H-extension modifies the supervisor mode to an \textit{hypervisor-extended supervisor mode} (HS-mode), which similarly to Intel's VT-x root mode, is orthogonal to the new \textit{virtual supervisor mode} (VS-mode) and \textit{virtual user mode} (VU-mode), and therefore can easily accommodate both bare-metal and hosted (a.k.a. type-1 and -2) as well as hybrid hypervisor architectures. 
Unavoidably, the extension also introduces two-stage address translations where the hypervisor controls the page tables mapping guest-physical to host-physical addresses. 
The \textit{virtualization mode} is controlled by the implicit V bit. When V is set, either VU- or VS-mode are executing and 2nd stage translation is in effect. When V is low, the system may execute in M-, HS- or U- mode. The extension defines a few new hypervisor instructions and CSRs, as well as extends existing machine CSRs to control the guest virtual memory and execution. For example, the \textit{hstatus} allows the hypervisor to track and control virtual machine exception behavior, \textit{hgatp} points to the 2nd-stage root page table, and the \textit{hfence} instructions allow the invalidation of any TLB entries related to guest translations. 
Additionally, it introduces a set of virtual supervisor CSRs which act as shadows for the original supervisor registers when the V bit is set. A hypervisor executing in HS-mode can directly access these registers to inspect the virtual machine state and easily perform context switches. The original supervisor registers which are not banked must be manually managed by the hypervisor.

An innovative mechanism introduced with the H-extension is the hypervisor virtual machine load and store instructions. These instructions allow the hypervisor to directly access the guest virtual address space to inspect guest memory without explicitly mapping it in its own address space. Furthermore, because these accesses are subject to the same permissions checks as normal guest access, it precludes against confused deputy attacks when, for example, accessing indirect hypercall arguments. This capability can be extended to user mode, by setting \textit{hstatus} {\color{blue}hypervisor user mode (HU)} bit. This further simplifies the implementation of (i) type-2 hypervisors such as KVM \cite{Dall2014}, which hosts device back-ends in userland QEMU, or (ii) microkernel-based hypervisors such as seL4 \cite{Klein2009}, which implement virtual machine monitors (VMMs) as user-space applications.

With regard to interrupts, in RISC-V, there are three basic types: external, timer, and software (essentially used as IPIs). Each interrupt can be re-directed to one of the privileged modes by setting the bit for target interrupt/mode in a per-hart (hardware-thread, essentially a core in RISC-V terminology) interrupt pending bitmap, which might be directly driven by some hardware device or set by software. This bitmap is fully visible to M-mode through the \textit{mip} CSR, while the S-mode has a filtered view of its interrupt status through \textit{sip}. This concept was extended to the new virtual modes through the \textit{hvip}, \textit{hip}, and \textit{vsip} CSRs. As further detailed in Section \ref{imp_hyp_enh}, in current RISC-V implementations, a hardware module called the CLINT (core-local interrupter) drives the timer and software interrupts, but only for machine mode.  Supervisor software must configure timer interrupts and issue IPIs via SBI, invoked through \textit{ecall} (environment calls, i.e., system call) instructions. The firmware in M-mode is then expected to inject these interrupts through the interrupt bitmap in the supervisor. The same is true regarding VS interrupts as the hypervisor must itself service guest SBI requests and inject these interrupts through \textit{hvip} while in the process invoking the machine-mode layer SBI. When the interrupt is triggered for the current hart, the process is inversed: (i) the interrupt traps to machine software, which must then (ii) inject the interrupt in HS-mode through the interrupt pending bitmap, and then (iii) inject it in VS mode by setting the corresponding \textit{hvip} bit. 
As for external interrupts, these are driven by the platform-level interrupt controller (PLIC) targeting both M and HS modes. The hypervisor extensions specifies a guest external interrupt mechanism which allows an external interrupt controller to directly drive the VS external interrupt pending bit. This allows an interrupt to be directly forward to a virtual machine without hypervisor intervention (albeit in an hypervisor controlled manner). However, this feature needs to be supported by the external interrupt controller. Unfortunately, the PLIC is not yet virtualization-aware. A hypervisor must fully trap-and-emulate PLIC accesses by the guest and manually drive the VS external interrupt pending bit in \textit{hvip}. The sheer number of traps involved in these processes is bound to impact interrupt latency, jitter, and, depending on an OS tick frequency, overall performance. For these reasons, interrupt virtualization support is one of the most pressing open-issues in RISC-V virtualization. In section \ref{imp_hyp_enh}, we describe our approach to address this issue.

The hypervisor extension also augments the trap encoding scheme with multiple exceptions to support VS execution. For instance, it adds guest-specific page faults exceptions for when translations at the second stage MMU fail, as well as VS-level ecalls which equate to hypercalls.

Finally, it is worth mentioning that the specification is tailored to seamlessly support nested virtualization; however, nested virtualization is out-of-scope of this article.


\section{Rocket Core Hypervisor Extension}
\label{imp_hyp_ext}

\begin{figure}[t]
    \centering
        \includegraphics[width=0.47\textwidth,clip]{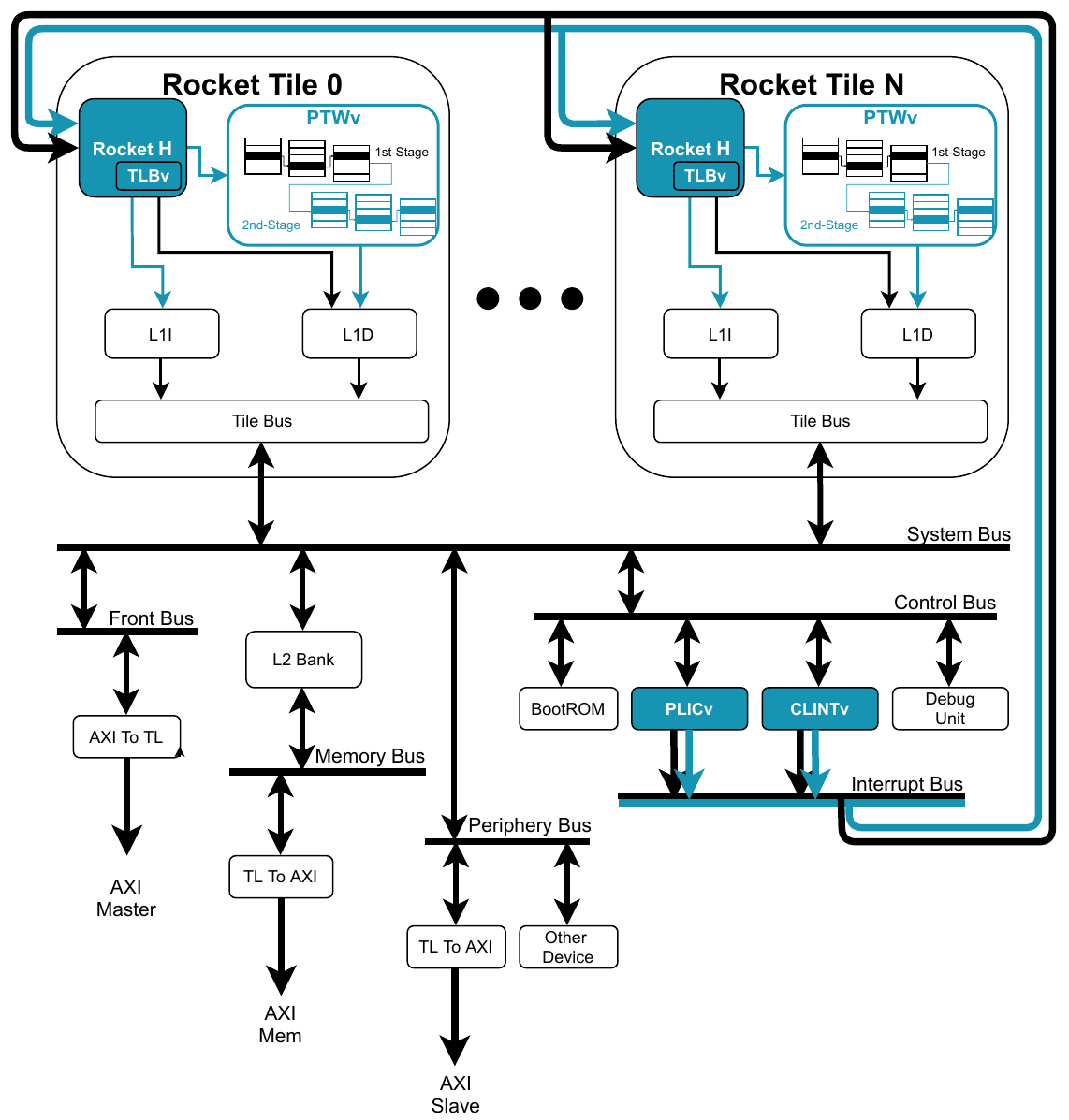}
    \caption{{\color{blue}Rocket chip diagram. H-extension and interrupt virtualization enhancements (CLINTv and PLICv) highlighted in blue. Adapted from \cite{Asanovic2016}.}}.
    \label{fig:rocket_chip_over}
\end{figure}

We have implemented the RISC-V hypervisor extension in the open-source Rocket core, a modern 5-stage, in-order, highly configurable core, part of the Rocket chip SoC generator \cite{Asanovic2016} and written in the novel Chisel hardware construction language \cite{bachrach2012} {\color{blue}(Figure \ref{fig:rocket_chip_over})}. Despite being possible to configure this core according to the 32- or 64-bit variants of the ISA (RV32 or RV64, respectively), our implementation currently only supports the latter. The extension can be easily enabled by adding a \textit{WithHyp} configuration fragment in a typical Rocket chip configuration. 

\mypara{{\color{blue}Hypervisor CSRs and Instructions.}} The bulk of our H-extension implementation in the Rocket core revolves around the CSR module which implements most of the privilege architecture logic: exception triggering and delegation, mode changes, privilege instruction and CSRs, their accesses, and respective permission checks. These mechanisms were straightforward to implement, as very similar ones already exist for other privilege modes. {\color{blue} Although most of the new CSRs and respective functionality mandated by the specification were implemented, we have left out some optional features. Specifically:

\begin{enumerate}
\item \textit{htimedelta} is a register that contains the difference between the value obtained by a guest when reading the \textit{time} register and the actual \textit{time} value. We expect this register to be emulated by the firmware running in M-mode as it is the only mode allowed to access \textit{time} (see CLINT background in section \ref{sec:timer});
\item \textit{htinst} and \textit{mtinst} are hardwired to zero. These registers expose a trapping instruction in an easy and pre-decoded form so that the hypervisor can quickly handle the trap while avoiding reading the actual guest instruction and polluting the data cache;
\item \textit{hgatp} is the hypervisor's 2nd-stage root page-table pointer register. Besides the pointer it allows to specify a translation mode (essentially, page-table formats and address space size) and the VMID (virtual machine IDs, akin to 1st-stage ASIDs). The current implementation does not support VMIDs and only allows for the \textit{Sv39x4} translation mode;
\item the \textit{hfence} instructions, which are the hypervisor TLB synchronization or invalidation instructions, always invalidate the full TLB structures. However, the specification allows to provide specific virtual addresses and/or a VMID to selectively invalidate TLB entries.
\end{enumerate}

}

Nevertheless, all the mandatory H-extension features are implemented and, therefore, our implementation is fully compliant with the RISC-V H-extension specification. Table \ref{tab:h-feat} summarizes all the included and missing features.

\begin{table}[t]
\caption{Current state of Hypervisor Extension features implemented in the Rocket core: \CIRCLE \hspace{1pt} fully-implemented; \LEFTcircle \hspace{1pt} partially implemented; \Circle \hspace{1pt} not implemented.}
\center
\begin{tabular}{|c|l|c|}
\hline
\multirow{11}{*}{CSRs} 
 & hstatus/mstatus   &  \CIRCLE  \\ \cline{2-3}   
 & hideleg/hedeleg/mideleg &  \CIRCLE  \\ \cline{2-3}
 & hvip/hip/hie/mip/mie    &   \CIRCLE \\ \cline{2-3}  
 & hgeip/hgeie   &   \CIRCLE  \\ \cline{2-3}         
 & hcounteren   &   \CIRCLE \\ \cline{2-3}  
 & htimedelta   &  \Circle \\ \cline{2-3}  
 & mtval2/htval   &   \CIRCLE  \\ \cline{2-3}        
 & mtinst/htinst   & \Circle   \\ \cline{2-3}  
 & hgapt   &  \LEFTcircle  \\ \cline{2-3}          
 &  \begin{tabular}[c]{@{}l@{}}vsstatus/vsip/vsie/vstvec/vsscratch\\vsepc/vscause/vstval/vsatp \end{tabular} &  \CIRCLE \\ \hline \hline   
\multirow{2}{*}{Intructions}
    & hlv/hlvx/hsv & \CIRCLE \\ \cline{2-3} 
    & hfence.vvma/gvma & \LEFTcircle \\ \hline \hline
\multirow{6}{*}{Exceptions \& Interrupts}
    & Environment call from VS-mode & \CIRCLE  \\ \cline{2-3} 
    & Instruction/Load/Store guest-page fault & \CIRCLE  \\ \cline{2-3} 
    & Virtual instruction &  \CIRCLE \\ \cline{2-3} 
    & \begin{tabular}[c]{@{}l@{}} Virtual Supervisor sw/timer/external\\interrupts\end{tabular} & \CIRCLE  \\ \cline{2-3} 
    & Supervisor guest external interrupt &  \CIRCLE \\ \hline
\end{tabular}
\label{tab:h-feat}
\end{table}

\begin{figure*}[t]
    \centering
    \includegraphics[height=7cm,width=1\textwidth,clip]{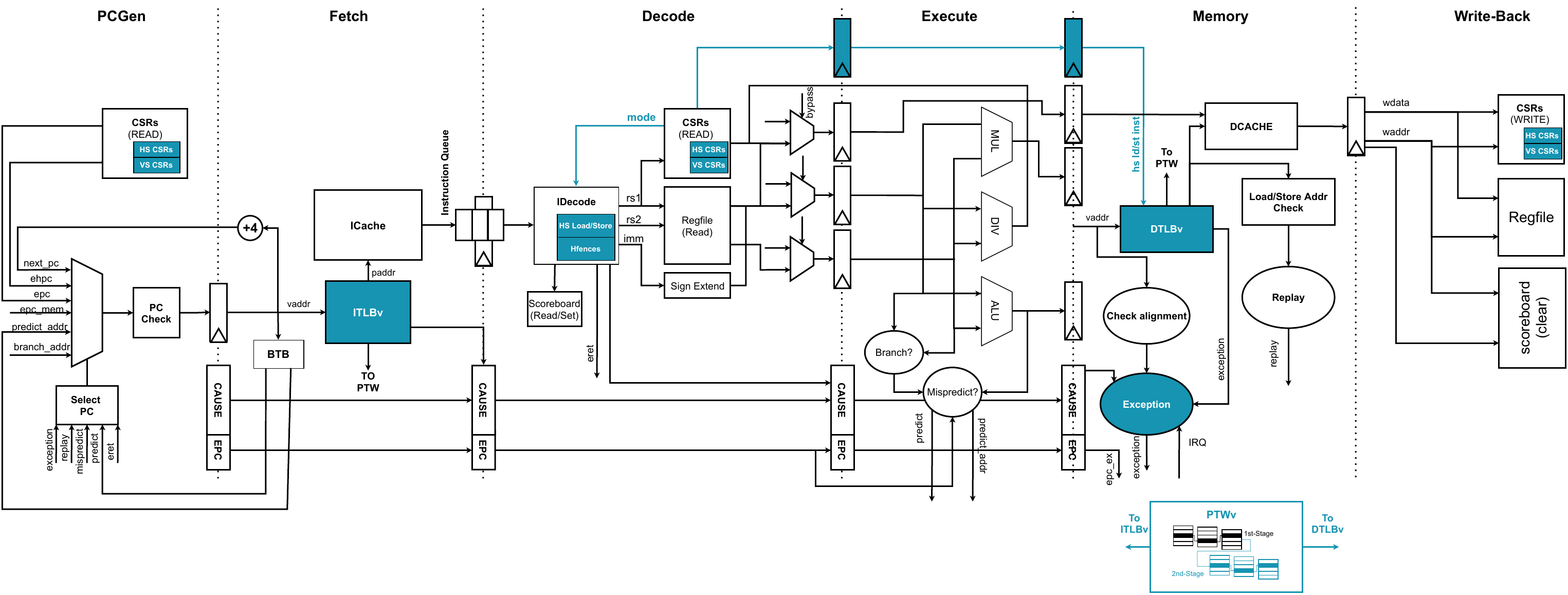}
   \caption{{\color{blue}Rocket core microarchitecture overview featuring the H-extension. Major architectural changes to the Rocket core functional blocks (e.g., Decoder, PTW, TLBs, and CSRs) are highlighted in blue. Adapted from \protect\footnotemark.}}
    \label{fig:rocket_core_micro}
\end{figure*}

\begin{figure}[t]
    \centering
        \includegraphics[width=0.47\textwidth,clip]{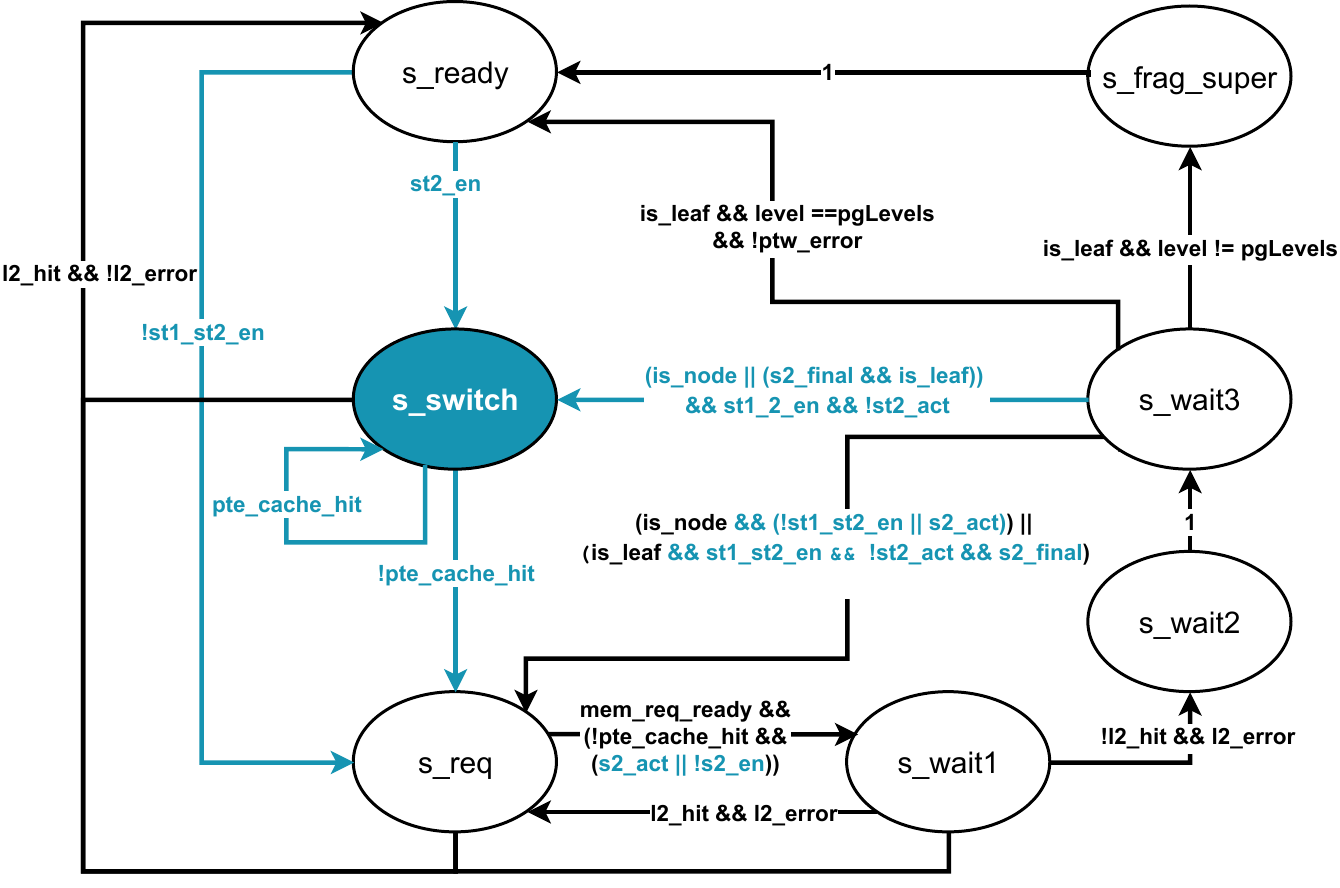}
    \caption{{\color{blue}PTW state machine featuring the two-stage translation. Modifications to the state machine, including a new state (\textit{s\_switch}) to switch the translation between the two stages (e.g., change the root page-table pointer), are highlighted in blue.}}
    \label{fig:two_st_machine}
\end{figure}

\mypara{{\color{blue}Two-stage Address Translation.}} The next largest effort focused on the MMU structures, specifically the page table walker (PTW) and translation-lookaside buffer (TLB), in particular, to add to support for the 2nd-stage translation. The implementation only supports the \textit{Bare} translation mode (i.e., no translation) and the \textit{Sv39x4}, which defines a specific page table size and topology which results in guest-physical addresses with a maximum width of 41-bits. The modification to the PTW extends the module's state-machine so that it switches to perform 2nd-stage translation at each level of the 1st translation stage {\color{blue}(Figure \ref{fig:two_st_machine})}. At each step it merges the results of both stages. When a guest leaf PTE (page table entry) is reached, it performs a final translation of the targeted guest-physical address. This proved to be one of the trickiest mechanisms to implement, given the large number of corner cases that arise when combining different page sizes at each level and of exceptions that might occur at each step of the process. TLB entries were also extended to store both the direct guest-virtual to host-physical address as well as the resulting guest-physical address of the translation. This is needed because even for a valid cached 2-stage translation, later accesses might violate one of the RWX permissions, and the specification mandates that the guest-physical address must be reported in \textit{htval} when the resulting exception is triggered. 
Note that the implementation does not support VMID TLB entry tagging. We have decided to neglect this optional feature for two mains reasons. Firstly, at the time of this witting, the Rocket core did not even support ASIDs. Secondly, static partitioning hypervisors (our main use case) do not use it at all. A different hypervisor must invalidate these structures at each context-switch. As such, the implemented support for \textit{hfence} instructions ignores the VMID argument. Furthermore, they invalidate all cached TLB or walk-cache entries used in guest translation, despite it specifying a virtual address argument, or being targeted at only the first stage (\textit{hfence.hvma}) or both stages (\textit{hfence.gvma}). To this end, an extra bit was added to TLB entries to differentiate between the hypervisor and virtual-supervisor translations. Finally, we have not implemented any optimizations such as dedicated 2nd-stage TLBs as many modern comparable processors do, which still leaves room for important optimizations.

\mypara{{\color{blue}Hypervisor Virtual-Machine Load and Store Instructions.}} Despite most of the implementation being straightforward, because it mainly involved replicating or extending existing or very similar functionality and mechanisms, the most invasive implemented feature was the support for hypervisor virtual-machine load and store instructions. This is because, although RISC-V already provided mechanisms for a privilege level to perform memory accesses subject to the same translation and restrictions of a lower privilege (such as by setting the {\color{blue}modify privilege - MPRV -} bit in \textit{mstatus}), these sync the pipeline instruction stream which results in the actual access permission modifications being associated with the overall hart state and not tagged to a specific instruction. As such, we added dedicated signals that needed to be propagated throughout the pipeline {\color{blue} starting from the decode stage, going through the L1 data cache up to the data TLB to signal the memory access originates from a hypervisor load/store}. {\color{blue} If this signal is set, the TLB will ignore the current access privilege (either HS or HU), and fetch the translation and perform the privilege checks as the access was coming from a virtual machine (VS or VU)}. Similar signals already existed for the fence instructions.

\mypara{{\color{blue}Other modifications.}} The Rocket chip generators truncate physical address width signals to the maximum needed for the configured physical address space. Thus, another issue we faced was the need to adapt the bit-width of some buses and register assumed to be carrying physical addresses to support guest-physical addresses, essentially virtual addresses. Finally, we also needed to slightly modify the main core pipeline to correctly forward virtual exceptions, including the new virtual instruction exception, to the CSR module along with other hypervisor-targeted information such as faulting guest-physical addresses to set the \textit{htval/mtval} registers.

\section{{\color{blue}Interrupt Virtualization Enhancements}} \label{imp_hyp_enh}

{\color{blue}As explained in the previous sections, RISC-V support for virtualization still only focuses on CPU virtualization. Therefore, as illustrated in Figure \ref{fig:rocket_chip_over}, we have also extended other Rocket chip components, namely the PLIC and the CLINT, to tackle some of the previously identified drawbacks regarding interrupt virtualization.}

\subsection{{\color{blue}Timer virtualization}}
\label{sec:timer}
\mypara{{\color{blue}CLINT background.}} {\color{blue} CLINT is the core-level interrupt controller responsible for maintaining machine-level software and timer interrupts in the majority of RISC-V systems. To inject software interrupts (IPIs in RISC-V lingo) in M-mode, the CLINT facilitates a memory-mapped register, denoted \textit{msip}, where each register is directly connected to a running CPU. Moreover, the CLINT also implements the RISC-V timer M-mode specification, more specifically the \textit{mtime} and \textit{mtimecmp} memory-mapped control registers}. \textit{mtime} is a free-running counter and a machine timer interrupt is triggered when its value is greater than the one programmed in \textit{mtimecmp}. There is also a read-only \textit{time} CSR accessible to all privilege modes, which is not supposed to be implemented but converted to a MMIO access of \textit{mtime} or emulated by firmware. Thus, M-mode software implementing the SBI interface (e.g., OpenSBI) must facilitate timer services to lower privileges via ecalls, by multiplexing logical timers onto the M-mode physical timer.

\mypara{{\color{blue}CLINT virtualization overhead.}} Naturally, this mechanism introduces additional burdens and impacts the overall system performance for HS-mode and VS-Mode execution, especially in high-frequency tick OSes.
As explained in section \ref{riscv_virt_supp}, a single S-mode timer event involves several M-mode traps, i.e., first to set up the timer and then to inject the interrupt in S-mode. This issue is further aggravated in virtualized environments as it adds extra HS-mode traps. The simplest solution to mitigate this problem encompasses providing multiple independent hardware timers directly available to the lower privilege levels, HS and VS, through new registers analogous to the M-mode timer registers. This approach is followed in other well-established computing architectures. For instance, the Armv8-A architecture has separate timer registers across all privilege levels and security states.

\begin{figure}[t]
    \centering
        \includegraphics[width=0.35\textwidth,clip]{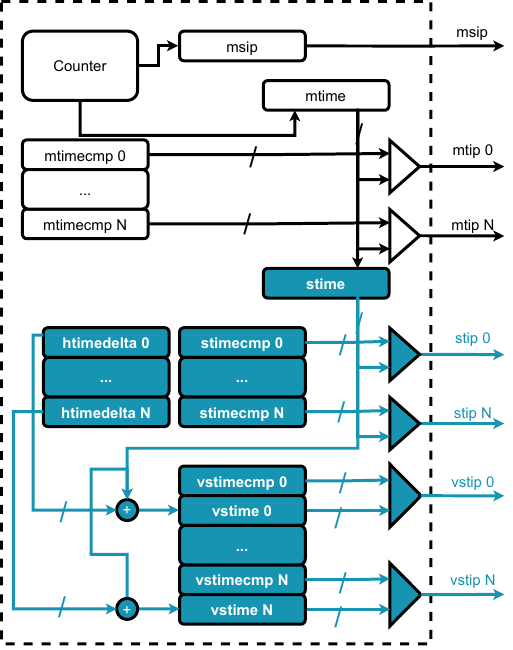}
    \caption{{\color{blue}CLINT microarchitecture with virtualization enhancements. Architectural changes to include hardware support for S and VS mode timers are highlighted in blue.}}
    \label{fig:clintv_over}
\end{figure}

\mypara{{\color{blue}CLINT virtualization extensions.}} As detailed in Table \ref{tbl_clint}, we added read-only \textit{stime} and \textit{vstime}, as well as read/write \textit{stimecmp} and \textit{vstimecmp} memory-mapped registers to the CLINT. Furthermore, we implemented a memory-mapped version of the \textit{htimedelta} CSR, which defines a drift of \textit{time} as viewed from VS- or VU-mode perspectives. 
In our implementation ({\color{blue} see Figure \ref{fig:clintv_over}}), \textit{htimedelta} will be reflected in the value of \textit{vstime}, by adding it to the value of \textit{mtime}. 
Adopting such approach would enable S-mode software to directly interact with its timer and receive interrupts without firmware mediation. In CLINT implementations each type of timer registers are mapped onto separate pages. However, hart replicas of the same register are packed contiguously in the same page. As such, with this approach, the hypervisor still needs to mediate VS- register access as it cannot isolate VS registers of each individual virtual hart (or vhart) using virtual memory. Nevertheless, traps from HS- to M-mode are no longer required, and when the HS and VS timer expires, the interrupt pending bit of the respective privilege level is directly set.

\begin{table}[t]
\captionof{table}{CLINT memory map. In bold, the new HS and VS timer registers.}
\centering
\begin{tabular}{l|l|l}
\hline
\textbf{Field}                   & \textbf{Offset(hex)}                & \textbf{Description}                                                                                                                                                         \\ \hline \hline
msip $\textit{n}$                & 0x00000 + ($\textit{n}$*4)          & M-mode hart $\textit{n}$ software interrupt                                                                                                                                  \\ \hline
mtimecmp $\textit{n}$            & 0x04000 + ($\textit{n}$*8)          & \begin{tabular}[c]{@{}l@{}}Compare value for the M-mode \\ hart $\textit{n}$ timer\end{tabular}                                                                              \\ \hline
mtime                            & 0x0bff8                             & Current time value                                                                                                                                                          \\ \hline
\textbf{stimecmp $\textit{n}$}   & \textbf{0x0c000 + ($\textit{n}$*8)} & \textbf{\begin{tabular}[c]{@{}l@{}}Compare value for the S\\ or HS-mode hart $\textit{n}$ timer\end{tabular}}                                                                \\ \hline
\textbf{stime}                   & \textbf{0x1bff8}                    & \textbf{\begin{tabular}[c]{@{}l@{}}Current time value for S-mode\\ (RO replica of mtime)\end{tabular}}                                                                       \\ \hline
\textbf{vstimecmp $\textit{n}$}  & \textbf{0x1c000 + ($\textit{n}$*8)} & \textbf{\begin{tabular}[c]{@{}l@{}}Compare value for the \\ VS-mode hart $\textit{n}$ timer\end{tabular}}                                                                    \\ \hline
\textbf{vstime $\textit{n}$}     & \textbf{0x14000 + ($\textit{n}$*8)} & \textbf{\begin{tabular}[c]{@{}l@{}}Current time value for the \\ VS-mode hart $\textit{n}$\\ (mtime + htimedelta $\textit{n}$)\end{tabular}}                                 \\ \hline
\textbf{htimedelta $\textit{n}$} & \textbf{0x24000 + ($\textit{n}$*8)} & \textbf{\begin{tabular}[c]{@{}l@{}}Holds the current time delta \\ between the value of the time \\ CSR and the value returned \\ in VS-mode hart $\textit{n}$\end{tabular}} \\ \hline
\end{tabular}
\label{tbl_clint}
\end{table}

\mypara{{\color{blue}CLINTv vs Other proposals.}} Concurrently to our work, there have been some proposals discussed among the community to include dedicated timer CSRs for HS and VS modes. The latest one which is officially under consideration, at a high-level, is very similar to our implementation. However, there are differences with regard to: firstly, it does not include \textit{stime} and \textit{vstime} registers but only the respective timer compares; secondly, and more important, we add the new timer registers as memory-mapped IO (MMIO) in the CLINT, and not as CSRs. The rationale behind our decision is based on the fact that the RISC-V specification states that the original M-mode timer registers are memory-mapped, due to the need to share them between all harts as well as due to power and clock domain crossing concerns. As the new timers still directly depend on the original \textit{mtime} source value, we believe its simpler to implement them as MMIO, centralizing all the timer logic. Otherwise, every hart would have to continuously be aware of the global \textit{mtime} value, possibly through a dedicated bus. Alternatively, it would be possible to provide the new registers, as well as \textit{htimedelta}, through the CSR interface following the same approach as the one used for \textit{time}, i.e., by converting the CSR accesses to memory accesses. This approach would, however, in our view, add unnecessary complexity to the pipeline as supervisor software can always be informed of the platform's CLINT physical address through standard methods (e.g., device tree).

\footnotetext{\url{https://inst.eecs.berkeley.edu/~cs250/fa13/handouts/lab2-riscv.pdf}}

\subsection{PLIC virtualization}
\label{sec:plic}
\begin{figure}[t]
    \centering
        \includegraphics[height=7cm,width=0.4\textwidth,clip]{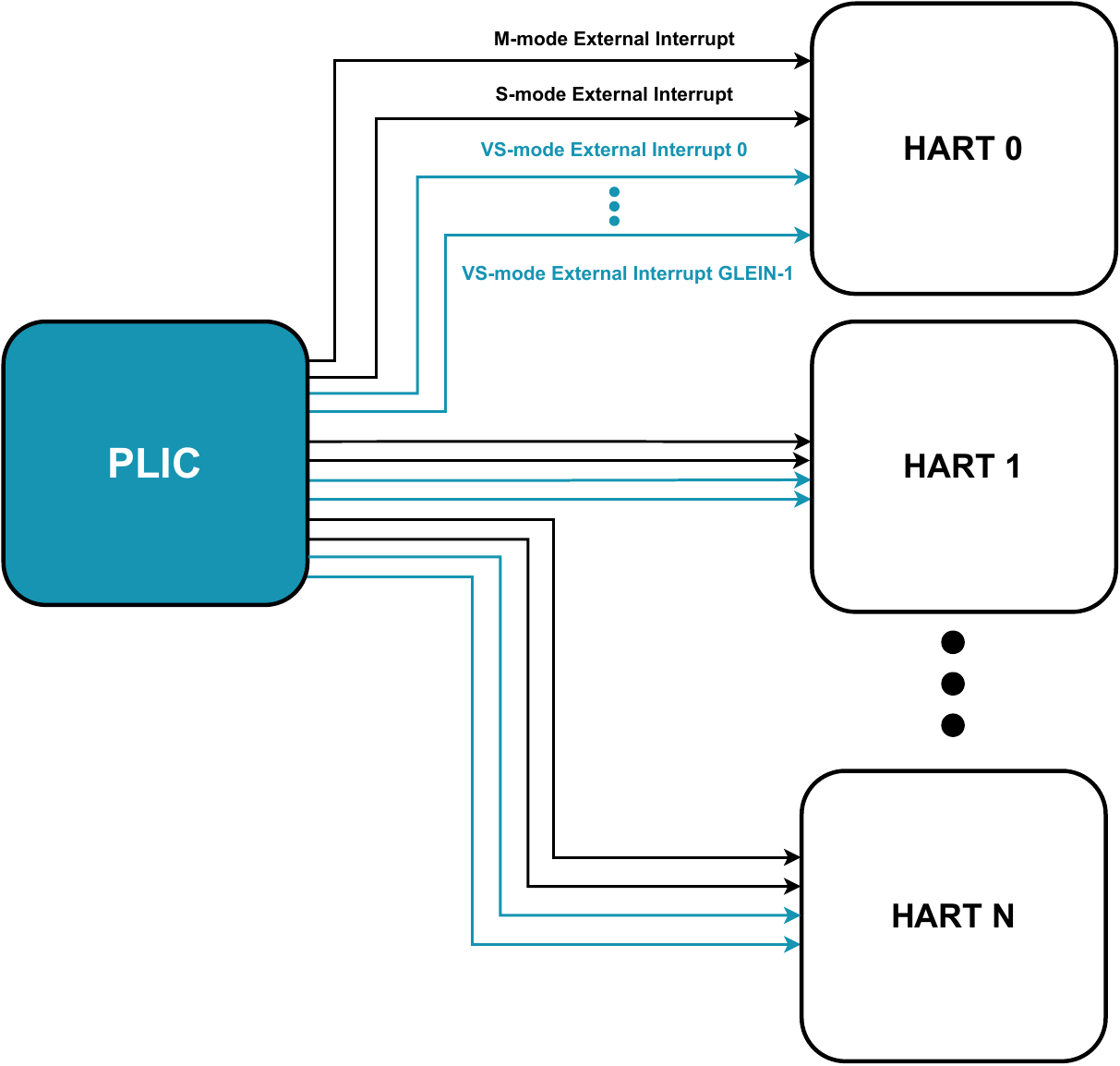}
    \caption{{\color{blue}High-level virtualization-aware PLIC logic.}}
    \label{fig:plicv}
\end{figure}

\mypara{{\color{blue}PLIC background.}} The PLIC is the external interrupt controller used in most current RISC-V systems. The PLIC is capable of multiplexing multiple devices interrupts to one or more hart contexts. More specifically, up to 1023 devices interrupt lines can be connected to the PLIC, each with its configurable priority. PLIC contexts represent a set of registers and external interrupts lines, each targeting a specific privilege level within a given hart (see Fig. \ref{fig:plicv}). Each line will drive the corresponding privilege bit in the hart global external interrupt pending bitmap. Part of each context, the PLIC provides registers for interrupt enabling as well as for interrupt handling: upon entry on an interrupt handler, software reads a context claim register which returns the interrupt ID that triggered the interrupt. To complete the interrupt handling process, the hart must write back to the \textit{complete register} the retrieved interrupt ID. The claimed interrupt will not be re-triggered until the interrupt during this process. Beyond that, PLIC also supports context interrupt masking through the \textit{threshold register}, i.e., interrupts with priority lower than the threshold value are not delivered to the context. Importantly, each set of claim/complete/threshold registers is mapped onto a different physical memory page.
 
\mypara{{\color{blue}PLIC virtualization overhead.}} Currently, only M-mode and S-mode contexts are supported (grey lines in Fig. \ref{fig:plicv}), meaning the PLIC specification does not provide additional interrupt virtualization support. The hypervisor is then responsible for emulating PLIC control registers accesses and fully managing interrupt injection into VS-mode. Emulating PLIC interrupt configuration registers, such as \textit{enable} and \textit{priority registers}, may not be critical as it is often a one-time-only operation performed during OS initialization. However, the same does not apply to the \textit{claim/complete registers}, which must be accessed before and after every interrupt handler. For physical interrupts directly assigned to guests, this is further aggravated, since it incurs in the extra trap to the hypervisor to receive the actual interrupt before injecting it in the virtual PLIC. These additional mode-crosses causes a drastic increase in the interrupt latency and might seriously impact overall system performance, especially for real-time systems that rely on low and deterministic latencies. 

\mypara{{\color{blue}PLIC virtualization enhancements.}} Based on the aforementioned premises, we propose a virtualization extension to the PLIC specification\footnote{https://github.com/josecm/riscv-plic-spec/tree/virt} that could significantly improve the system's performance and latency by leveraging the guest external interrupt feature of the RISC-V hypervisor extension (see section \ref{riscv_virt_supp}). We had four main requirements: (i) allow direct assignment and injection of physical interrupts to the active VS-mode hart, i.e., without hypervisor intervention; (ii) minimize traps to the hypervisor, in particular, by eliminating traps on claim/complete register access; (iii) allow a mix of purely virtual and physical interrupts for a given VM; and (iv) a minimal design with a limited amount of additional hardware cost and low overall complexity. We started by adding \textit{GEILEN} VS-mode contexts to every hart. \textit{GEILEN} is a hypervisor specification "macro" that defines the maximum (up to 64) available VS external interrupt contexts that can be simultaneously active for a hart, in a given implementation. This is a configurable parameter in our implementation. The context for the currently executing vhart is selected through the VGEIN field in the \textit{hstatus} CSR. Fig. \ref{fig:plicv} highlights, in blue, the external interrupt lines associated with VS contexts that directly drive the hart's VS mode external interrupt pending bits. With additional virtual contexts and independent context's \textit{claim/complete register} pair available on separate pages, the hypervisor can allow direct guest access to \textit{claim/complete registers} by mapping them in the guests' physical address space, mitigating the need to trap-and-emulate such registers. However, access to configuration registers such as \textit{priority} and \textit{enable registers} are still trapped since these registers are not often accessed and are not in the critical path of interrupt handling. A hypervisor might assign physical interrupts to a guest as follows. When a guest sets configurations register fields for a given interrupt, a hypervisor commits it to hardware in the vhart's context if the interrupt was assigned to its VM. Otherwise, it might save it in memory structures if it corresponds to an existing virtual interrupt fully managed by the hypervisor.

\begin{figure}[t]
    \centering
        \includegraphics[width=0.46\textwidth,clip]{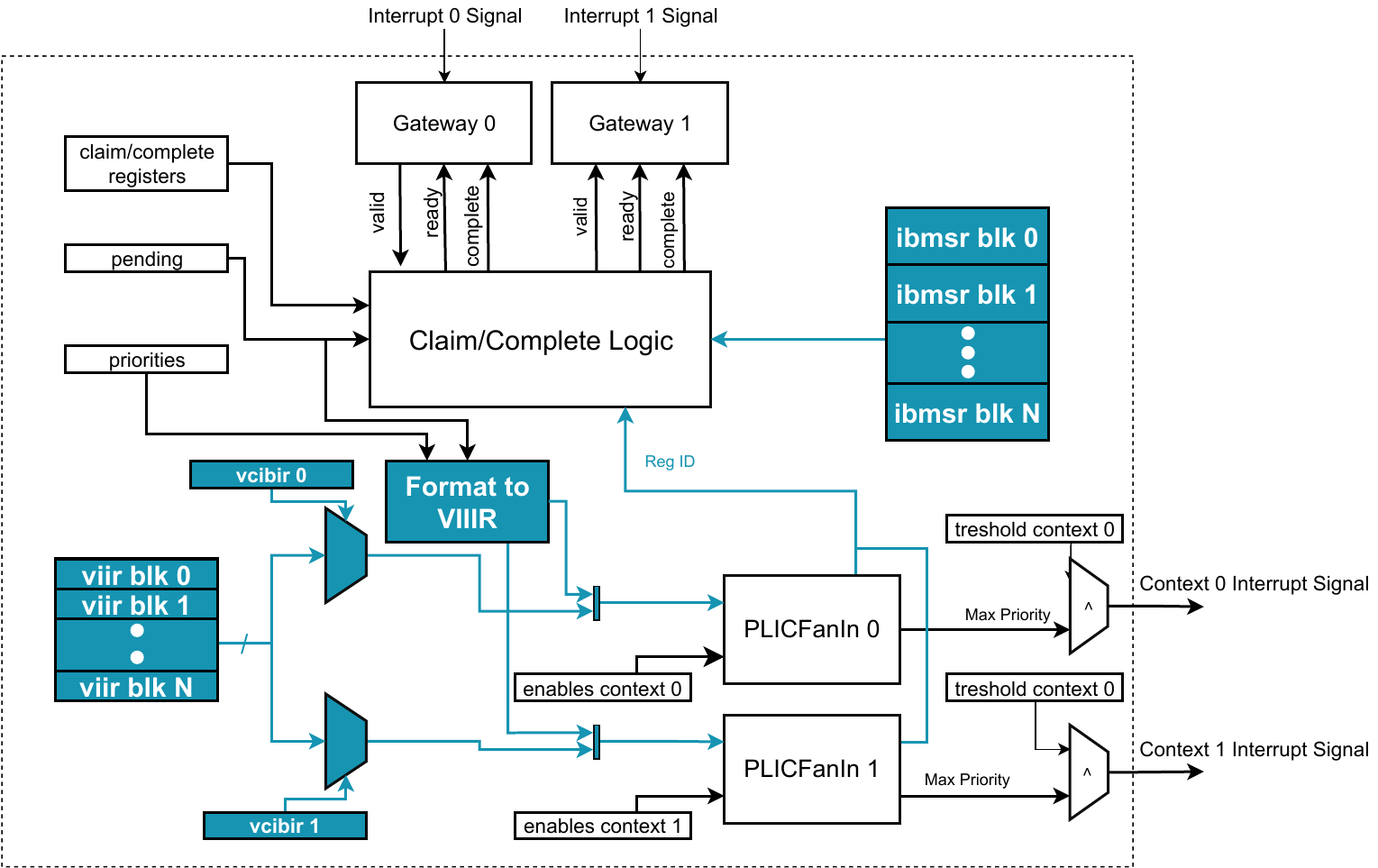}
    \caption{{\color{blue}PLIC microarchitecture with virtualization enhancements. Architectural changes to include support for virtual interrupt injection are highlighted in blue.}}
    \label{fig:plicv_micro}
\end{figure}


\begin{table}[t]
\captionof{table}{Extended memory map for the virtualization-aware PLIC. In bold, the new virtual interrupt injection registers.}
\centering
\begin{tabular}{l|l|l}
\hline
\textbf{Field}                                & \textbf{Offset}                                                                                                   & \textbf{Description}                                                                                                                        \\ \hline \hline
priority $\textit{i}$                         & \begin{tabular}[c]{@{}l@{}}0x0000000\\ ($\textit{i}$ * 4)\end{tabular}                                            & Interrupt ID $\textit{i}$ priority register                                                                                                 \\ \hline
pending                                       & 0x0001000                                                                                                         & \begin{tabular}[c]{@{}l@{}}Interrupt source pending bits.\\ Up to 32 sources per register.\end{tabular}                                      \\ \hline
enable $\textit{c}$                           & \begin{tabular}[c]{@{}l@{}}0x0002000 + \\ ($\textit{c}$ * 0x80)\end{tabular}                                      & \begin{tabular}[c]{@{}l@{}}Interrupt source enable registers \\ bits for context $\textit{c}$. Upto 32\\ sources per register.\end{tabular} \\ \hline
threshold $\textit{c}$                        & \begin{tabular}[c]{@{}l@{}}0x0200000 + \\ ($\textit{c}$ * 0x1000)\end{tabular}                                    & \begin{tabular}[c]{@{}l@{}}Priority threshold register for \\ context $\textit{c}$\end{tabular}                                             \\ \hline
claim/complete $\textit{c}$                   & \begin{tabular}[c]{@{}l@{}}0x0200004 + \\ ($\textit{c}$ * 0x1000)\end{tabular}                                    & \begin{tabular}[c]{@{}l@{}}Claim/Complete register for \\ context $\textit{c}$\end{tabular}                                                 \\ \hline
\textbf{vcibir $\textit{c}$}                  & \textbf{\begin{tabular}[c]{@{}l@{}}0x4000000 + \\ ($\textit{c}$ * 4)\end{tabular}}                                & \textbf{\begin{tabular}[c]{@{}l@{}}Virtual context $\textit{c}$ injection\\  block ID register\end{tabular}}                                \\ \hline
\textbf{viir $\textit{j}$ block $\textit{n}$} & \textbf{\begin{tabular}[c]{@{}l@{}}0x4010000 + \\ ($\textit{n}$ * 0x1000) + \\ ($\textit{j}$ * 4)\end{tabular}} & \textbf{\begin{tabular}[c]{@{}l@{}}Virtual interrupt injection \\ register $\textit{j}$ of block $\textit{n}$\end{tabular}}                 \\ \hline
\textbf{ibmsr block $\textit{n}$}             & \textbf{\begin{tabular}[c]{@{}l@{}}0x4110000 + \\ ($\textit{n}$ * 4)\end{tabular}}                                & \textbf{\begin{tabular}[c]{@{}l@{}}Injection block management \\ and status register for block $\textit{n}$\end{tabular}}                   \\ \hline
\end{tabular}
\label{tlb:plicv_memmap}
\end{table}

\mypara{{\color{blue}PLICv pure virtual interrupts extensions.}} With direct access to \textit{claim/complete registers} at the guest level, injection of purely virtual interrupts must also be done through the PLIC, so there are unified and consistent forwarding and handling for all the vhart's interrupts. To this end, and inspired by Arm's GIC list registers, we added three new memory-mapped 32-bit wide register sets to the PLIC to support this operation (see Table \ref{tlb:plicv_memmap} {\color{blue}and Figure \ref{fig:plicv_micro}}): \textit{Virtual Interrupt Injection Registers} (\textit{VIIR}), \textit{Virtual Context Injection Block ID Registers} (\textit{VCIBIR}), and \textit{Injection Block Management and Status Registers} (\textit{IBMSR}). \textit{VIIRs} are grouped into page-sized \textit{injection blocks}.  The number of \textit{VIIRs} in a block and the number of available blocks are implementation-defined (up to a maximum of 1000 and 240, respectively); however, if a given block exists, at least one register must be implemented. There are \textit{GEILEN} \textit{VCIBIR} per-hart which are used to specify the source injection block used for a given context's virtual interrupt injection. In this way, virtual interrupts for multiple harts belonging to a specific VM can be injected through a single injection block, precluding the need for complex synchronization across hypervisor harts. Also, this allows a hart to directly inject an interrupt in a foreign vhart without forcing an extra HS trap. Setting a context's \textit{VCIBIR} to zero indicates that no injection block is attached.

\mypara{{\color{blue}PLICv virtual interrupt injection mechanism.}} Virtual interrupt injection is done through the \textit{VIIRs} which are composed of three fields: \textit{inFlight}, \textit{interruptID}, and \textit{priority}. 
Setting the VIIR with a \textit{interruptID} greater than 0 and the \textit{inFlight} bit not set would make the interrupt pending for the virtual contexts associated with its block. The bit \textit{inFlight} is automatically set when the virtual interrupt is pending and a claim is performed indicating that the interrupt is active, preventing the virtual interrupt from being pending. When the \textit{complete register} is written with an ID present in a VIIR, that register is cleared, otherwise an interrupt might be raised to signal it, as explained next. On a \textit{claim register} read, the PLIC selects the higher priority pending interrupt of either a context's injection block or enabled physical interrupts.
Each block is associated with a \textit{block management interrupt} (akin to GIC's maintenance interrupt) fed back through the PLIC itself with implementation-defined IDs. It serves to signal events related to the block's \textit{VIIRs} lifecycle. Currently, there are two well-defined events: (i) no \textit{VIIR} pending, and (ii) claim write of an non-present \textit{interruptID}. The enabling of each type of event, signaling of currently pending events and complementary information are done through a corresponding \textit{IBMSR}. 
Note that all of the new PLIC registers are optional as the minimum number of injection blocks is zero. If this is the case, a hypervisor might support either (i) a VM with only purely virtual interrupts, falling back to the full trap-and-emulate model, or (ii) a VM with only physical interrupts. 



\mypara{{\color{blue}PLICv context switch.}} An important point regarding the PLIC virtualization extensions we have somewhat neglected in our design is its impact on vhart context-switch. At first sight, it might seem prohibitively costly due to the high number of MMIO context and block registers to be saved and restored. However we believe this is minimized first (i) due to the possibility of having up to \textit{GEILEN} virtual contexts for each hart and a number of \textit{injection blocks} larger than the maximum possible number of active vharts; second (ii) we expect that only a small number (1 to 4) VIIRs are implemented in each block; and third (iii) as we expect that physical interrupt assignment will be sparse for a given  virtual machine, an hypervisor  can  keep  a  word-length bitmap of the enable registers that contain these interrupts, and only save/restore the absolutely needed.


\section{Bao RISC-V Porting} \label{bao_port}

\mypara{{\color{blue}Bao in a nutshell.}} Bao \cite{Martins2020} is an open-source static partitioning hypervisor developed with the main goal of facilitating the straightforward consolidation of mixed-criticality systems, thus focusing on providing strong safety and security guarantees. It comprises only a minimal, thin-layer of privileged software leveraging ISA virtualization support to partition the hardware, including 1-to-1 virtual to physical CPU pinning, static memory allocation, and device/interrupt direct assignment. Bao implements a clean-slate, standalone component featuring about 8 KSLoC (source lines of code), which depends only on standard firmware to initialize the system and perform platform-specific tasks such as power management. It provides minimal inter-VM communication facilities through statically configured shared memory and notifications in the form of interrupts. It also implements from the get-go simple hardware partitioning mechanisms such as cache coloring to avoid interference in caches shared among the multiple VMs. 

\mypara{{\color{blue}Bao for RISC-V.}} Bao originally targeted only Arm-based systems. So, we have initially ported to RISC-V using the QEMU implementation of the H-extension. Later, this port was used in the Rocket core with the extensions described in this article without any friction. Given the simplicity of both the hypervisor and RISC-V designs, in particular, the virtualization extensions {\color{blue} and the high degree of similarity with Arm's Aarch64 ISA}, the port was mostly effortless. It comprised an extra 3731 SLoC compared to Arm's 5392 lines of arch-specific code. Nevertheless, a small step-back arose that forced the need to modify the hypervisor's virtual address space mappings. For security reasons, Bao used the recursive page table mapping technique so that each CPU would map only the memory it needs and not map any internal VM structures besides its own or any guest memory. RISC-V impose some constrains given that each PTE must either serve has a next-table pointer or a leaf PTE. Therefore, we had to modify Bao to identity-map all physical memory, to enable performing software page walks to build guest page tables. {\color{blue} Another RISC-V-specific facility we had to implement in Bao was the read guest memory operation. This need arose because neither QEMU nor our Rocket implementation of the H-extension provides the pre-decoded trapped instruction on \textit{htinst}. Therefore, on a guest trap due to a page-fault that requires emulation (e.g., PLIC access), Bao must read the instruction from guest memory and decode it in software. Nevertheless, Bao never directly maps guest memory in its own address space. It reads the instruction via hypervisor load instructions.} 
The RISC-V Bao port relies on a basic SBI firmware implementation featuring the IPI, Timer, and RFENCE extensions. As of this writing, only OpenSBI has been used. Bao provides SBI services to guests so these can program timer interrupts and send IPI between vharts. However, Bao mostly acts as a shim for most VS- SBI calls, as the arguments are just processed and/or sanitized and the actual operation is relegated to the firmware SBI running in machine mode.

\mypara{\color{blue} Bao RISC-V Limitations} It is also worth mentioning there are still some gaps in the ISA to suors like Bao. {\color{blue} For example, cache maintenance support operations are not standardized.  At the moment, core implementations must provide this functionality via custom facilities. The Rocket core provides a custom machine-mode instruction for flushing the L1 data cache. As Bao relies on these operations to implement some of its features (e.g., cache coloring), we have implemented a custom cache flush API in OpenSBI that Bao calls when it needs to clean caches. Another issue regards external interrupt support. Due to the PLIC's virtualization limitations (see section \ref{sec:plic}), Bao's implementation must fully trap and emulate guest access to the PLIC, i.e., not only on configuration but also on interrupt delivery and processing. As we show in section \ref{sec:eval-int-lat}, this adds significant overheads, especially on interrupt latency.} Finally, Bao relies on IOMMU support to be able to directly assign DMA-capable devices to guest OSes. However, there is no standard IOMMU currently available in RISC-V platforms (see section \ref{discussion}).


\section{Evaluation} \label{eval}

The evaluation was conducted for three different SoC configurations, i.e., dual-, quad-, and six-core Rocket chip with per-core 16 KiB L1 data and instruction caches, and a shared unified 512 KiB L2 LLC (last-level cache). The software stack encompasses the OpenSBI (version 0.9), Bao (version 0.1), and Linux (version 5.9), and bare metal VMs. OpenSBI, Bao, and bare metal VMs were compiled using the GNU RISC-V Toolchain (version 8.3.0 2020.04.0), with -O2 optimizations. Linux was compiled using the GNU RISC-V Linux Toolchain (version 9.3.0 2020.02-2). Our evaluation focused on functional verification (Section \ref{eval-fval}),  hardware resources (Section \ref{eval-hw-over}), performance and inter-VM interference (Section \ref{eval-perf-int}), and interrupt latency (Section \ref{eval-hw-over}). 



\subsection{Functional Verification} \label{eval-fval}

The functional verification of our hardware implementation was performed on a Verilator-generated simulator and on a Zynq UltraScale+ MPSoC ZCU104 FPGA. 

We have developed an ad-hoc testing framework as a bare-metal application. Our goal was to test individual features of the hypervisor specification without any additional system software complexity and following a test-driven development (TDD) like approach. During development, we have written a comprehensive test suite spanning features such as two-stage translation, exception delegation, virtual exceptions, hypervisor load-store instructions, CSR access, just to name a few. To be able to test out individual features, the framework provides an API which easily allows: (i) fully resetting processor state at beginning of each test unit; (ii) fluidly and transparently changing privilege mode; (iii) easy access a guest virtual address with any combination of 1st and 2nd stage permissions; and (iv) easy detection and recovery of exceptions and later checking of its state and causes. Nevertheless, the framework still has some limitations such as not allowing user mode execution or experimenting with superpages. This hypervisor extension testing framework and accompanying test suite are openly available\footnote{https://github.com/josecm/riscv-hyp-tests} and can be easily adapted to other platforms. We have also run our test suite in QEMU, unveiling bugs in the hypervisor extension implementation, for which patches were later submitted.

As a second step to validate our implementation we have successfully ran two open-source hypervisors: Bao and XVisor \cite{Patel2015}. XVisor also provides a "Nested MMU Test-suite" which mainly exercises the two-stage translation. At the time of this writing, our implementation fully passes this test suite. Some bugs uncovered while running these hypervisors were translated into tests and incorporated into our test suite. 

\subsection{Hardware Overhead} \label{eval-hw-over}

\begin{table}[t]
\caption{Rocket chip hardware resource overhead with virtualization extensions}
\centering
\begin{tabular}{l|l|l|l|l}
\hline
\multicolumn{2}{l|}{}                                                                   & \textbf{Dual-Core}         & \textbf{Quad-Core}          & \textbf{Six-Core}        \\ \hline  \hline
\multirow{2}{*}{\textbf{\begin{tabular}[c]{@{}l@{}}Rocket\\ Cores\end{tabular}}} & LUTs & $\sfrac{50922}{\textbf{11\%}}$     & $\sfrac{101744}{\textbf{12\%}}$    & $\sfrac{152957}{\textbf{12\%}}$ \\ \cline{2-5} 
                                                                                 & Regs & $\sfrac{25086}{\textbf{30\%}}$     & $\sfrac{50172}{\textbf{30\%}}$           & $\sfrac{75258}{\textbf{30\%}}$        \\ \hline
\multirow{2}{*}{\textbf{CLINT}}                                                  & LUTs & $\sfrac{68}{\textbf{375\%}}$          & $\sfrac{196}{\textbf{296\%}}$          & $\sfrac{269}{\textbf{373\%}}$       \\ \cline{2-5} 
                                                                                 & Regs & $\sfrac{194}{\textbf{297\%}}$         & $\sfrac{324}{\textbf{336\%}}$          & $\sfrac{454}{\textbf{277\%}}$       \\ \hline
\multirow{2}{*}{\textbf{PLIC}}                                                   & LUTs & $\sfrac{90}{\textbf{140\%}}$ & $\sfrac{144}{\textbf{236\%}}$ & $\sfrac{220}{\textbf{263\%}}$       \\ \cline{2-5} 
                                                                                 & Regs & $\sfrac{83}{\textbf{325\%}}$          & $\sfrac{116}{\textbf{412\%}}$          & $\sfrac{149}{\textbf{460\%}}$       \\ \hline
\multirow{2}{*}{\textbf{Others}}                                                 & LUTs & $\sfrac{11207}{\textbf{2\%}}$      & $\sfrac{13242}{\textbf{3\%}}$       & $\sfrac{91821}{\textbf{0,5\%}}$   \\ \cline{2-5} 
                                                                                 & Regs & $\sfrac{4257}{\textbf{0,1\%}}$       & $\sfrac{4628}{\textbf{0,2\%}}$        & $\sfrac{4728}{\textbf{2\%}}$     \\ \hline
\multirow{2}{*}{\textbf{Total}}                                                  & LUTs & $\sfrac{62287}{\textbf{11\%}}$     & $\sfrac{115356}{\textbf{11\%}}$    & $\sfrac{167753}{\textbf{11\%}}$ \\ \cline{2-5} 
                                                                                 & Regs & $\sfrac{29620}{\textbf{27\%}}$     & $\sfrac{55250}{\textbf{28\%}}$      & $\sfrac{80589}{\textbf{29\%}}$  \\ \hline
\end{tabular}
\label{ref:tbl_hard}
\end{table}

To assess the hardware overhead, we synthesized multiple SoC configurations with an increasing number of harts (2, 4, and 6). We used Vivado 2018.3 targeting the Zynq UltraScale+ MPSoC ZCU104 FPGA. Table \ref{ref:tbl_hard} presents the post-synthesis results, depicting the number of look-up tables (LUTs) and registers for the three SoC configurations. For each cell, there is the absolute value for the target configuration and, in bold, the relative increment (percentage) compared to the same configuration with the hypervisor extensions disabled. We withhold data on other resources (e.g., BRAMs or DSPs) as the impact on its usage is insignificant. 

According to Table \ref{ref:tbl_hard}, we can draw two main conclusions. First, there is an overall non-negligible cost to implement the hypervisor extensions support: an extra 11\% LUTs, and 27-29\% registers. Diving deeper, we observed that this overhead comes almost exclusively from two sources: the CSR and TLB modules. The CSR increase is explained given the number of HS and VS registers added by the H-extension specification. The increase in the TLB is mainly due to the widening of data store to hold guest-physical addresses (see section \ref{imp_hyp_ext}) and the extra privilege-level and permission match and check complexity. The second important point is that, although the enhancements to the CLINT and PLIC reflect a large relative overhead, as these components are simple and small compared to the overall SoC infrastructure, there is no significant impact on the total hardware resources cost.
Lastly, we can also highlight that increasing the number of available harts in the SoC does not impact the relative hardware costs.



\begin{figure*}[t]
    \centering
    \includegraphics[height=5cm,width=1\textwidth,clip]{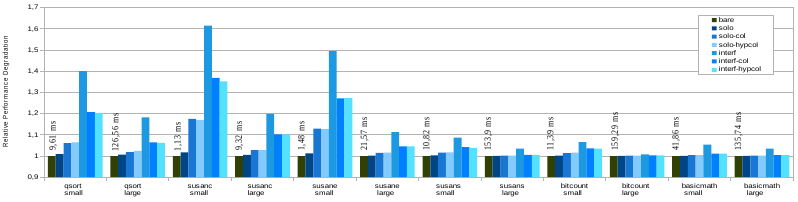}
   \caption{Relative performance overhead of MiBench automotive suite for different system configurations, relative to bare-metal execution. Absolute value indicated at the top of the solo bar.}
    \label{fig:mibench}
\end{figure*}

\subsection{Performance and Inter-VM Interference} \label{eval-perf-int}

To assess performance overhead and inter-hart / inter-VM interference, we used the MiBench Embedded Benchmark Suite. MiBench is a set of 35 benchmarks divided into six suites, each one targeting a specific area of the embedded market. We focus our evaluation on the automotive subset. The automotive suite includes three high memory-intensive benchmarks, i.e., more susceptible to interference due to LLC and memory contention (qsort, susan corners, and susan edges). 

Each benchmark was ran for seven different system configurations targeting a six-core design: (i) guest native execution (bare); (ii) hosted execution (solo); (iii) hosted execution with cache coloring for VMs (solo-col); (iv) hosted execution with cache coloring for VMs and the hypervisor (solo-hypcol); (v) hosted execution under interference from multiple colocated VMs (interf); (vi) hosted execution under interference with cache coloring for VMs (interf-col); and (vii) hosted execution under interference with cache coloring for VMs and the hypervisor (interf-hypcol). Hosted scenarios with cache partitioning aim at evaluating the effects of partitioning micro-architectural resources at the VM and hypervisor level and to what extent it can mitigate interference. We execute the target benchmark in a Linux-based VM running in one core, and we add interference by running one VM pinned to five harts, each running a bare-metal application. Each hart runs an ad-hoc bare-metal application that continuously writes and reads a 1 MiB array with a stride equal to the cache line size (64 bytes). The platform's cache topology allows for 16 colors, each color consisting of 32 KiB. When enabling coloring, we assign seven colors (224 KiB) to each VM. The remaining two colors were reserved for the hypervisor coloring case scenario.

The experiments were conducted in Firesim, an FPGA-accelerated cycle-accurate simulator, deployed on an AWS EC2 F1 instance, running with a 3.2 GHz simulation clock. 
Fig. \ref{fig:mibench} presents the results as performance normalized to bare execution, meaning that higher values report worse results. Each bar represents the average value of 100 samples. For each benchmark, we added the execution time (i.e., absolute performance) at the top of the bare-metal execution bar.

According to Fig. \ref{fig:mibench}, we can draw six main conclusions. Firstly, hosted execution (solo) causes a marginal decrease of performance (i.e., average 1\% overhead increase) due to the virtualization overheads of 2-stage address translation. Secondly, when coloring (solo-col and solo-hypcol) is enabled, the performance overhead is further increased. This extra overhead is explained by the fact that only about half of the L2 cache is available for the target VM, and that coloring precludes the use of superpages, significantly increasing TLB pressure. Thirdly, when the system is under significant interference (inter), there is a considerable decrease of performance, in particular, for the memory-intensive benchmarks, i.e., qsort (small), susan corners (small), and susan edges (small). For instance, for the susan corners (small) benchmark, the performance overhead increases by 62\%. Fourthly, we can observe that cache coloring can reduce the interference (inter-col and inter-hypcol) by almost 50\%, with a slight advantage when the hypervisor is also colored. Fifthly, we can observe that the cache coloring, per se, is not a magic bullet for interference. Although the interference is reduced, it is not completely mitigated, because the performance overhead for the colored configurations under interference (inter-col and inter-hypcol) is different from the ones without interference (solo-col and solo-hypcol). Finally, we observe that the less memory-intensive benchmarks (i.e., basicmath and bitcount) are less vulnerable to cache interference and that benchmarks handling smaller datasets are more susceptible to interference.        


The achieved results for RISC-V share a similar pattern to the ones assessed for Arm \cite{Martins2020}. In our previous work \cite{Martins2020}, we have deeply investigated the micro-architectural events using a performance monitoring unit (PMU). As part of future work, we plan to conduct a deeper evaluation of micro-architectural interference effects while proposing additional mechanisms to help mitigate inter-VM interference.

\subsection{Interrupt Latency} \label{sec:eval-int-lat}
\begin{figure}[t]
    \centering
        \includegraphics[height=4cm,width=0.47\textwidth,clip]{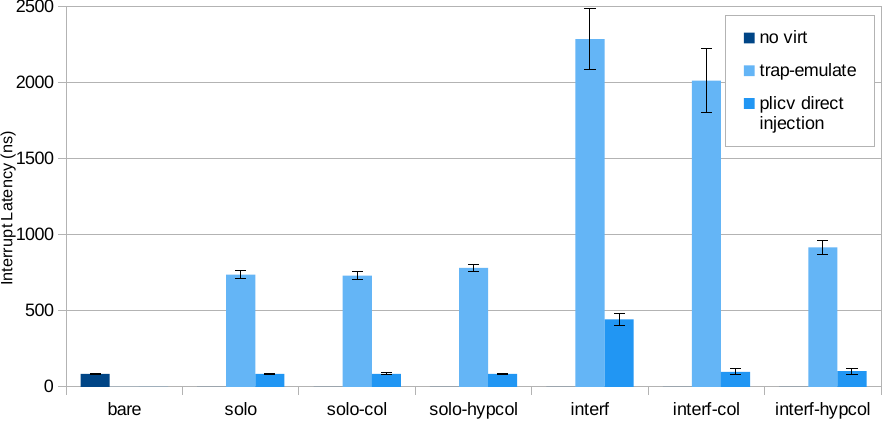}
    \caption{Interrupt latency in nanoseconds for bare-metal execution and for guest execution with injection following a trap-and-emulate approach or direct injection with hardware support.}
    \label{fig:int_lat}
\end{figure}

To measure interrupt latency and respective interference, we use a custom bare-metal benchmark application and a custom MMIO, auto-restart, incrementing timer that drives a PLIC interrupt input line. This application sets up the timer to trigger an interrupt at 100Hz (each 10 ms). The latency corresponds to the value read from the timer counter at the start of the interrupt handler. We invalidate the L1 instruction cache at each measurement using the \textit{fence.i} instruction as we believe it is more realistic to assume this cache is not hot with regard to the interrupt handling or the hypervisor's interrupt injection code. 
We ran the benchmark using Firesim with the same platform and configurations described in section \ref{eval-perf-int}. For guest configurations we took samples for both trap-and-emulate and PLIC interrupt direct injection. The average of the results obtained from 100 samples (the first 2 discarded) for each configuration are depicted in Fig. \ref{fig:int_lat}.

The interrupt latency for the bare (in Fig. \ref{fig:int_lat}, no virt) execution is quite low (approx. 80 ns) and steady. The trap-and-emulate approach introduces a penalty of an order of magnitude (740 ns) that is even more significant under interference (up to 2280 ns, about 300\%) both in average and standard deviation. Applying cache partitioning via coloring helps to mitigate this, which shows that most of the interference happens in the shared L2 LLC. The difference between inter-col and interf-hypcol shows that it is of utmost importance to assign dedicated cache partitions to the hypervisor: the interfering VM also interferes with the hypervisor while injecting the interrupt and not only with the benchmark code execution itself.

Fig. \ref{fig:int_lat} also shows that the effect of the direct injection achieved with guest external interrupt and PLIC virtualization support can bring guest interrupt latency to near-native values. Furthermore, it shows only a fractional increase under interference (when compared to the trap-and-emulate approach) which can also be attenuated with cache coloring. As the hypervisor no longer intervenes in interrupt injection, for this case, it suffices to color guest memory. A small note is that the use of cache coloring does not affect the benchmark for solo execution configurations, given that the benchmark code is very small. Thus, the L1 and L2 caches can easily fit both the benchmark and hypervisor's injection code. Finally, we can conclude that with PLIC virtualization support, it is possible to significantly improve external interrupt latencies for VMs. 



\section{Related Work} \label{relatedwork}

\par There is a broad spectrum of hardware virtualization related technologies and hypervisor solutions. Due to the extensive list of works in the literature, we will focus on (i) COTS and custom hardware virtualization technology and extensions and (ii) hypervisors and microkernels solutions for RISC-V. 

\mypara{Hardware virtualization technology.} Modern computing architectures such as x86 and Arm have been adding added hardware extensions to assist virtualization to their CPUs for more than a decade. Intel has developed the Intel Virtualization Technology (Intel VT-x) \cite{Uhlig2005}, the Advanced Programmable Interrupt Controller (APIC) virtualization extension (APICv), and Intel Virtualization Technology for Directed I/O (Intel VT-d). Intel has also included nested virtualization hardware-based capabilities with Virtual Machine Control Structure (VMCS) Shadowing. Arm included the virtualization extensions (Arm VE) since Armv7-A and developed additional hardware to the Generic Interrupt Controller (vGIC) for efficient virtual interrupt management. Recently, Arm has announced a set of extensions in the Armv8.4-A that includes the addition of secure virtualization support \cite{Arm2018sel2} and the Memory System Resource Partition and Monitoring (MPAM) \cite{arm2018mpam}. There are additional COTS hardware technologies that have been leveraged to assist virtualization, namely the MIPS virtualization module \cite{Moratelli2016}, AMD Virtualization (AMD-V), and Arm TrustZone \cite{Pinto2018, Pinto2019}. The academia has also been focused on devising and proposing custom hardware virtualization support and mechanisms \cite{Xu2017, Lim2017}. Xu et al. proposed vCAT \cite{Xu2017}, i.e., dynamic shared cache management for multi-core virtualization platforms based on Intel’s Cache Allocation Technology (CAT). With NEVE \cite{Lim2017}, Lim et al. developed a set of hardware enhancements to the Armv8.3-A architecture to improve nested virtualization performance, which was then included in Armv8.4-A. Within the RISC-V virtualization scope, Huawei has also presented extensions both to the CLINT and the PLIC \cite{Zhao2020}, including one of the timer extension proposals mentioned in section \ref{imp_hyp_enh}. Regarding the PLIC, comparing to our approach, their design is significantly more invasive and complex, as it uses memory-resident tables for interrupt/vhart and vhart/hart mappings. Their approach significantly complicates the PLIC implementation as it must become a bus-master. Nevertheless, this might bring some advantages, e.g., speeding-up VM context-switches. Furthermore, they also propose to extend the CLINT not only to include supervisor and virtual-supervisor timers but also to allow direct send and receive of HS/VS software interrupts (i.e., IPIs) without firmware/hypervisor intervention

\mypara{Hypervisors and microkernels for RISC-V.} KVM \cite{Lublin2007} and Xvisor \cite{Patel2015} were the first hypervisors adding support for the RISC-V H-extension in QEMU. KVM \cite{Lublin2007} is a type-2 hosted hypervisor integrated into Linux's mainline as of 2.6.20. KVM targets mainly enterprise virtualization setups for data centers and private clouds. Xvisor \cite{Patel2015} is a type-1 monolithic hypervisor targeting embedded systems with soft real-time requirements. Both hypervisors are officially part of the RISC-V Software Ecosystem\footnote{https://github.com/riscv/riscv-software-list} and naturally have been used by technical groups as reference implementations to validate and evolve the H-extension. RVirt\footnote{https://github.com/mit-pdos/RVirt} is an S-mode trap-and-emulate hypervisor for RISC-V, written in Rust. Contrarily to KVM and XVisor, RVirt can run in RISC-V processors without hardware virtualization support. Diosix\footnote{https://diosix.org/} is another lightweight bare-metal hypervisor written in Rust for RISC-V. Similar to RVirt, Diosix can run in RISC-V cores that lack the H-extension, leveraging the physical memory protection (PMP) to achieve isolation. Xtratum, a hypervisor primarily developed for safety-critical aerospace applications, has also recently been ported to support RISC-V ISA \cite{gomez2020risc}, following the same PMP-based concept for isolation as Diosix. Xen \cite{Hwang2008} and Jailhouse \cite{Ramsauer2017}, two widely used open-source hypervisor solutions, have already given preliminary steps towards RISC-V support. However, as of this writing, upstream support for RISC-V is not yet available, but it is expected to be included in the foreseeable future. seL4, a formally verified microkernel, is also verified on RISC-V \cite{Heiser2020}. Other commercial microkernels already support RISC-V. Preeminent examples include the SYSGO PikeOS and the Wind River VxWorks.


\section{Discussion} \label{discussion}

\mypara{\color{blue} Hypervisor Extension Specification State.} The RISC-V H-extension is currently in its 0.6.1 version and is being developed within the privileged specification working group of RISC-V International, following a well-defined extension development lifecycle. The specification draft has been stable for quite some time and therefore is approaching a frozen state, after which it will enter a period of public review before finally being ratified. However, to enter a frozen state it will need both (i) open RTL core implementations suitable for deployment as soft-cores on FPGA platforms and (ii) hypervisor ports that exercise its mechanisms and provide feedback. Until the extensions are ratified, we do not expect any commercial IP or ASIC implementations to be available. With this work, we have contributed with one open RTL implementation but more are needed. Xvisor and KVM have been the reference open-source hypervisors used in the extension development process. We have further contributed with the Bao port, but the more hypervisor ports are available to evaluate the suitability of the H-extension for different hypervisor architectures, the better.

\mypara{\color{blue} Missing Architectural and Virtualization Features.} As discussed in the article, there are still some gaps in RISC-V, in particular with respect to virtualization. At the ISA level, features like cache management operations are needed. Fortunately, there is already a working group defining these mechanisms. At a platform level, timer and external interrupt virtualization support is needed. Our results show the importance of these mechanisms to achieve low and deterministic interrupt latency in virtualized real-time systems. There are already efforts within the RISC-V community to provide this support: a new extension proposal is on the fast track to include dedicated timers for HS- VS-modes; and a new interrupt controller architecture featuring support for message-signaled interrupts (MSI) and virtualization support is under development within the privileged specification working group. Another missing component critical for virtualization is the IOMMU. An IOMMU is needed to implement efficient virtualization, by allowing the direct assignment of DMA-capable devices to VMs, while guaranteeing strong isolation between VMs and the hypervisor itself. Static partitioning hypervisors such as Bao completely depend on IOMMU, as they do not provide any kind of device emulation and only pass-through access. At the moment, in a RISC-V platform, a Bao guest that wishes to use a DMA device must have all its memory configured with identity mapping. Unfortunately, this still completely breaks encapsulation, serving only for experimentation and demonstration purposes, not being suitable for production.

\mypara{\color{blue} Multi-core Interference Mitigation.} In section \ref{eval}, we have demonstrated something well-understood and documented in the literature \cite{Yun2013, Mancuso2013, Kloda2019, Xu2019, Martins2020, Farshchi2020}, i.e., that (i) in multi-core platforms there is significant inter-core interference due to shared micro-architectural resources (e.g. caches, buses, memory controllers), (ii) which can be minimized by mechanisms such as page coloring used to partition shared caches. Other techniques such as memory bandwidth reservations \cite{Yun2013} and DRAM bank partitioning \cite{yun2014} can minimize interference further ahead in the memory hierarchy. These partitioning mechanisms are important in embedded mixed-criticality systems both from the security and safety perspectives by protecting against side-channel attacks and guaranteeing determinism and freedom-from-interference required by certification standards (e.g. ISO26262). They are also useful for server systems by helping to guarantee quality-of-service (QoS) and increase overall utilization \cite{Lo2015}. 
However, software-based approaches typically have significant overheads and increase the trusted computing base (TCB) complexity. Academic works such as Hybcache \cite{dessouky2020} or the bandwidth regulation unit (BRU) \cite{Farshchi2020} propose the implementation of this kind of mechanisms in RISC-V cores (Ariane \cite{zaruba2019cost} and Rocket respectively). {\color{blue} SafeSU \cite{Cabo2021} provides similar features by relying on a hardware statistics unit that measures inter-core interference in commercial space-graded RISC-V MPSoC.} SiFive has provided cache partitioning mechanisms in hardware via way-locking \cite{Gwennap2020}. We have started experimenting with these mechanisms using Bao and will present our findings in future work. During this work, we found it would be useful to have a standard set of mechanisms and interfaces to rely on. We argue that RISC-V is also missing a standard extension to provide such facilities. Other ISAs have already introduced these ideas, e.g., Intel's CAT and Arm's MPAM \cite{arm2018mpam}. MPAM functionality is also extended to other virtualization-critical system-bus masters including the GIC and the SMMU (Arm's interrupt controller and IOMMU, respectively), something that should also be taken into account when developing similar RISC-V specifications.

\mypara{\color{blue} Alternative Partitioning Approach.} Even without virtualization support, it is possible to implement static partitioning in RISC-V leveraging the trap-and-emulate features described in section \ref{riscv_virt_supp} and using the PMP for memory isolation instead of two-stage translation. The PMP is a RISC-V standard component that allows M-mode software to white-list physical address space regions on a per-core basis. This results in a kind of para-virtual approach, as the guest must be aware of the full physical address space and possibly recompiled for different system configurations. To provide direct assignment of DMA devices, the host platform would also need to provide IOPMPs (akin to IOMMU, without translation), which is a specification already on course. Furthermore, the hypervisor would be forced to flush micro-architectural state such as TLBs or virtual caches at each context switch resulting in significant performance overheads. The use of VMIDs, part of the H-extension, tackles this issue. Notwithstanding, this is not a real problem for statically partitioned systems. Thus, once there is no commercial hardware featuring the H-extension available in the market, this is the approach of some of the hypervisors mentioned in section \ref{relatedwork}. We are currently developing a customized version of Bao to run in RISC-V platforms without H-extension support (e.g., Microchip PolarFire SoC Icicle or the upcoming PicoRio). Nevertheless, we believe the hypervisor extension is still a better primitive for implementing these systems, given the higher flexibility and scalability it provides. 



\section{Conclusion}
\par In this article, we have presented the first implementation of the RISC-V H-extension in a real RISC-V core, i.e. Rocket core. We have also proposed a set of hardware enhancements to the interrupt controller and the timer infrastructure aiming at tackling mixed-criticality systems requirements of minimal latencies, determinism and predictability. To validate and evaluate our hardware implementation, we have also ported the Bao open-source hypervisor to RISC-V. We achieved functional verification of our implementation in a Verilator-generated simulator and a Zynq UltraScale+ MPSoC ZCU104 FPGA. We carried out an extensive set of experiments in FireSim, a cycle-accurate simulator, to assess performance, inter-VM interference, and interrupt latency. The results demonstrated that the H-extension, per se, introduces a reduced performance penalty, but without additional hardware support interference and interrupt latency can impose a prohibitive cost for MCSs. Our proposed architectural enhancements considerably minimize these effects, by reducing interrupt latency and interference by an order of magnitude. Lastly, we discussed identified gaps existing in RISC-V with regard to virtualization and we outlined internal ongoing efforts within RISC-V virtualization. Our hardware design was made freely available for the RISC-V community and is currently the single reference implementation available to ratify the H-extension specification. 



\section*{Acknowledgment}
This work has been supported by FCT -\textit{ Funda\c{c}\~ao para a Ci\^encia e Tecnologia} within the R\&D Units Project Scope: UIDB/00319/2020. This work has also been supported by FCT within the PhD Scholarship Project Scope: SFRH/BD/138660/2018.

\bibliographystyle{IEEEtran}
\bibliography{main}

\begin{IEEEbiography}
    [{\includegraphics[width=1in,height=1.25in,clip]{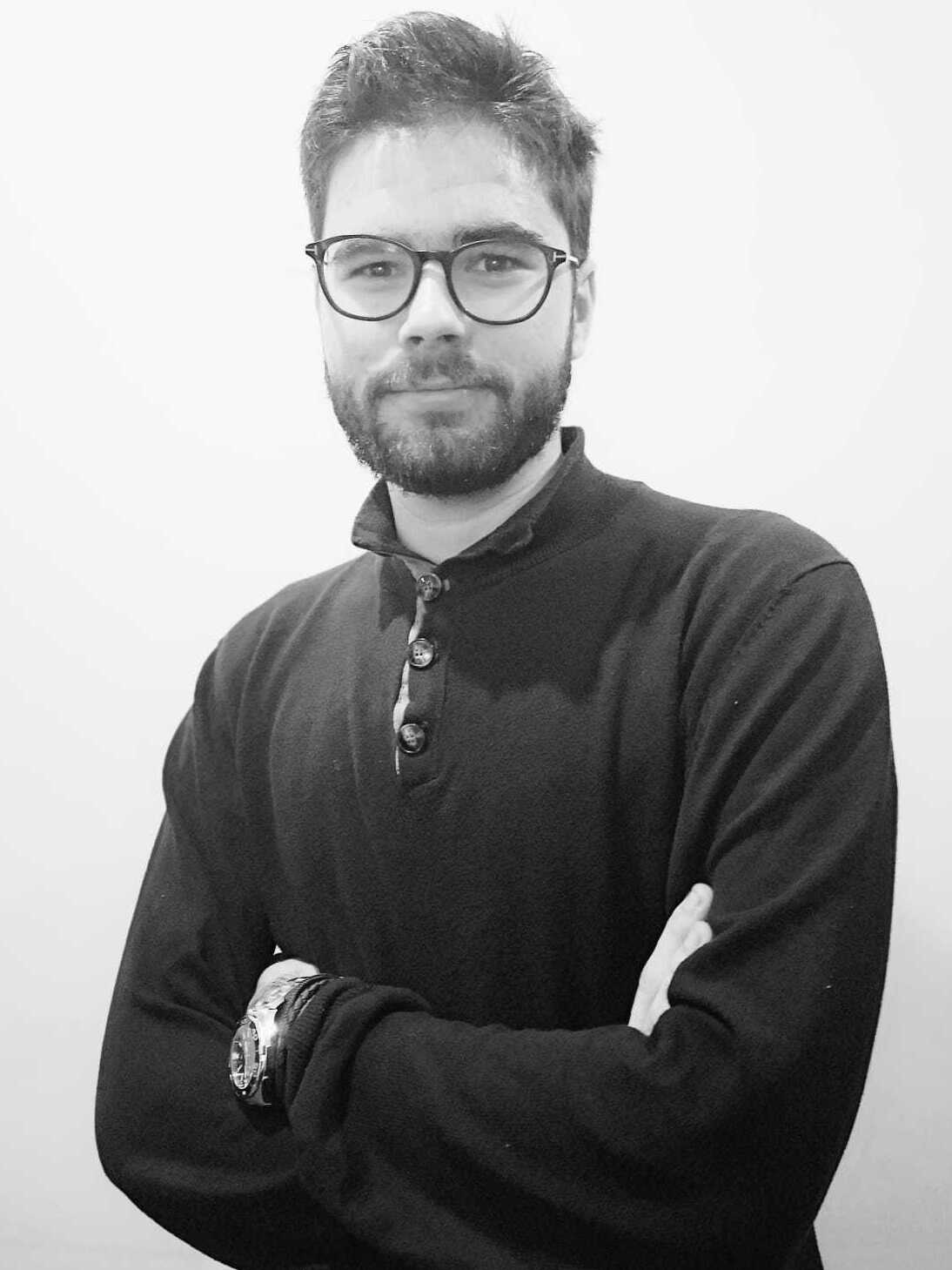}}]{Bruno Sá} is a Ph.D. student at the  Embedded  Systems  Research Group, University of Minho, Portugal. Bruno holds a Master in Electronics and Computer Engineering with specialization in Embedded Systems and Automation,Control and Robotics. Bruno's areas of interest include operating systems, virtualization for embedded systems, computer architectures, IoT systems, and artificial intelligence. Contact him at bruno.vilaca.sa@gmail.com.
\end{IEEEbiography}

\begin{IEEEbiography}
    [{\includegraphics[width=1in,height=1.25in,clip]{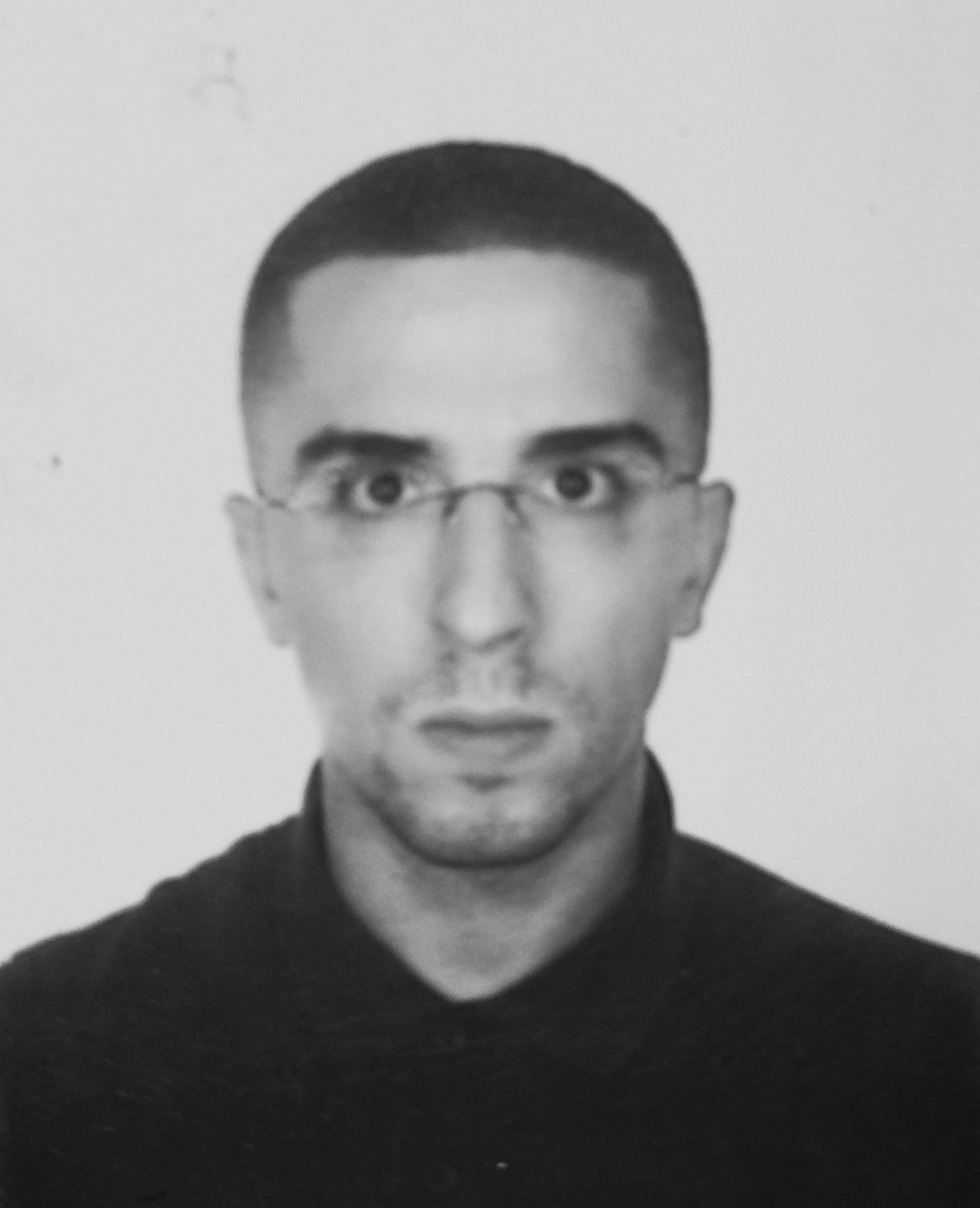}}]{José Martins} is a Ph.D. student and teaching assistant at the Embedded Systems Research Group, University of Minho, Portugal. José holds a Master in Electronics and Computer Engineering - during his Masters, he also was a visiting student at the University of Wurzburg, Germany. Jose has a significant background in operating systems and virtualization for embedded systems. Over the last few years, he has also been involved in several projects on the aerospace, automotive, and video industries. He is the main author of the Bao hypervisor. Contact him at jose.martins@dei.uminho.pt.
\end{IEEEbiography}

\begin{IEEEbiography}
    [{\includegraphics[width=1in,height=1.25in,clip]{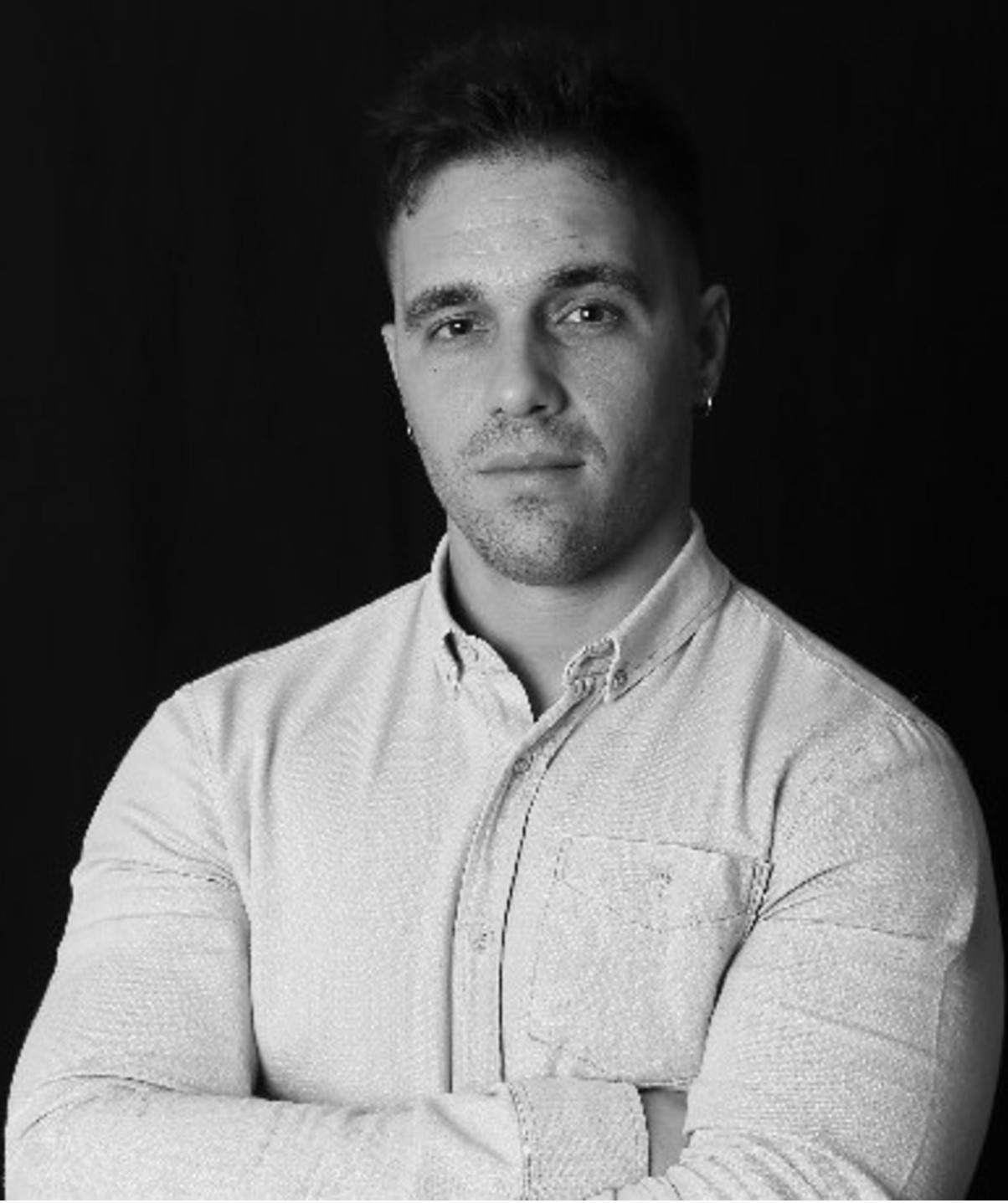}}]{Dr. Sandro Pinto} is a Research Scientist and Invited Professor at the University of Minho, Portugal. He holds a Ph.D. in Electronics and Computer Engineering. During his Ph.D., Sandro was a visiting researcher at the Asian Institute of Technology (Thailand), University of Wurzburg (Germany), and Jilin University (China). Sandro has a deep academic background and several years of industry collaboration focusing on operating systems, virtualization, and security for embedded, cyber-physical, and IoT-based systems. He has published several scientific papers in top-tier conferences/journals and is a skilled presenter with speaking experience in several academic and industrial conferences. Contact him at sandro.pinto@dei.uminho.pt.
\end{IEEEbiography}

\end{document}